\documentclass[]{aa}      
\usepackage{geometry}                       
\usepackage[varg]{txfonts}
\usepackage[utf8]{inputenc}
\usepackage{graphicx}               
\usepackage{amssymb}
\usepackage[english]{babel}
\usepackage{xcolor}
\usepackage{placeins}
\usepackage{afterpage}
\usepackage{booktabs}

\bibpunct{(}{)}{;}{a}{}{,} 

\title{The kinematic signature of the Galactic warp in Gaia DR1}
\subtitle{I. The \emph{Hipparcos} sub-sample}

\author{E. Poggio\inst{1,2}
\and R. Drimmel\inst{2} \and R. L. Smart\inst{2,3} \and A. Spagna\inst{2} \and M. G. Lattanzi\inst{2} } \institute{
Universit\`a di Torino, Dipartimento di Fisica, via P. Giuria 1, 10125 Torino, Italy
\and
Osservatorio Astrofisico di Torino, Istituto Nazionale di Astrofisica (INAF), Strada Osservatorio 20, 10025 Pino
Torinese, Italy
\and
School of Physics, Astronomy and Mathematics, University of Hertfordshire, College Lane, Hatfield AL10 9AB, UK\\
}

\date{ }  

\begin{document}
\abstract
{The mechanism responsible for the warp of our Galaxy, as well as its dynamical nature, continues to remain
unknown. With the advent of high precision astrometry, new horizons have been
opened for detecting the kinematics associated with the warp and constraining possible warp formation scenarios for the
Milky Way. }
{The aim of this contribution is to establish whether the first \emph{Gaia} data release (DR1) 
shows significant evidence of the kinematic signature expected from a long-lived Galactic warp in the kinematics of
distant OB stars. As the first paper in a series, we present our approach for analyzing the proper motions and apply it
to the sub-sample of \emph{Hipparcos} stars. }
{We select a sample of 989 distant spectroscopically-identified OB stars from the New Reduction of
\emph{Hipparcos} \citep{VanLeeuwen2007}, of which 758 are also in the first \emph{Gaia} data release (DR1),
covering distances from 0.5 to 3 kpc from the Sun. We
develop a model of the spatial distribution and kinematics of the OB stars from which we produce
 the probability distribution functions of the proper motions, with and without the systematic motions expected from a long-lived warp. A
likelihood analysis is used to compare the expectations of the models with the observed proper motions from both
\emph{Hipparcos} and \emph{Gaia} DR1.
 }
{ We find that the proper motions of the nearby
OB stars are consistent with the signature of a kinematic warp, while those of the more distant stars (parallax $<$ 1 mas) are not. 
 }
{ The kinematics of our sample of young OB stars suggests that systematic vertical motions in the disk cannot be
explained by a simple model of a stable long-lived warp. The warp of the Milky Way may either be a transient feature,
or additional phenomena are acting on the gaseous component of the Milky Way, causing systematic vertical motions that
are masking the expected warp signal. A larger and deeper sample of stars with \emph{Gaia} astrometry will be needed
to  constrain the dynamical nature of the Galactic warp. }


\maketitle

\section{ Introduction }

It has been known since the early HI 21-cm radio surveys that the outer gaseous disk of the Milky Way is warped with
respect to its flat inner disk \citep{Burke:1957,Kerr:1957,Westerhout:1957,Oort:1958}, bending upward in the north (I
and II Galactic quadrants) and downward in the south (III and IV Galactic quadrants). The Galactic warp has since been
seen in the dust and stars \citep{Freudenreich:1994, Drimmel:2001,
LopezCorredoira:2002B,Momany:2006,Marshall:2006,Robin:2008,Reyle:2009}.
Our Galaxy is not peculiar with respect to other disk galaxies: more than 50 percent of spiral galaxies are warped 
\citep{SanchezSaveedra:1990,Reshetnikov:1998,Guijarro:2010}. The high occurrence of
warps, even in isolated galaxies, implies that either these features are easily and continuously generated, or that
they are stable over long periods of time. In any case, the nature and origin of the galactic warps in general are
still unclear \citep{Sellwood:2013}.

While many possible mechanisms for generating warps in disk galaxies have been proposed, which is actually at work for
our own Galaxy remains a mystery. This is due to the fact that while the shape of the Galactic warp is known, its
dynamical nature is not; vertical systematic motions associated with the warp are not evident in radio surveys that
only reveal the velocity component along
the line-of-sight. Being located within the disk of the Milky Way, systematic vertical motions will primarily manifest
themselves to us in the direction perpendicular to our line-of-sight. More recent studies of the neutral HI component
\citep{Levine:2006,Kalberla:2007} confirm that the Galactic warp is already evident at a
galactocentric radius of 10 kpc, while the warp in the dust and stellar components are observed to start inside or very
close to the Solar circle \citep{Drimmel:2001,Derriere:2001,Robin:2008}. Thus, if the warp is stable, the associated
vertical motions should be evident in the component of the stellar proper motions perpendicular to the Galactic disk.

A first attempt to detect a kinematic signature of the warp in the proper motions of stars was first made using OB
stars \citep{Miyamoto:1988}. 
More recently, a study of the kinematic warp was carried out by \cite{Bobylev:2010,Bobylev:2013}, 
claiming a connection between the stellar-gaseous warp and the kinematics of their tracers, 
namely nearby red clump giants from Tycho-2 and cepheids with UCAC4 proper motions.
Using red clump stars from the PPMXL survey, \cite{LopezCorredoira:2014}
concluded that the data might be consistent with a long-lived warp, 
 though they admit that smaller systematic errors in
the proper motions are needed to confirm this tentative finding. Indeed, large-scale systematic errors in the
ground-based proper motions compromise efforts to detect the Galactic warp. The first real hopes of overcoming such
systematics came with global space-based astrometry. However, using \emph{Hipparcos} \citep{Hip:1997} data for OB stars,
\cite{Smart:1998} and \cite{Drimmel:2000} found that the kinematics were consistent neither with a warp nor with a flat
unwarped disk.

Before the recent arrival of the first {\emph Gaia} Data Release \citep[DR1]{Brown:2016}, the best
all-sky astrometric accuracy is found in the New Reduction of the \emph{Hipparcos} catalogue \citep[HIP2]{VanLeeuwen2007},
which improved the quality of astrometric data by more than a factor of two with respect to the original \emph{Hipparcos}
catalogue. For the \emph{Hipparcos} subsample the {\emph Gaia} DR1 astrometry is improved further by more than an order of
magnitude. The primary aim of this work is to assess whether either the HIP2 or the new \emph{Gaia} astrometry for the OB
stars in the \emph{Hipparcos} shows any evidence of the systematics expected from a long-lived warp. 
We choose the OB stars as they are intrinsically bright, thus can be seen to large distances,
and are short-lived, so are expected to trace the motions of the gas from which they were born. 
We select stars with spectral types of B3 and earlier. For B3 stars, stellar evolutionary models \citep[e.g.][]{Chen:2015} 
predict masses ranging from 7 to 9 $M_{\odot}$, corresponding to a MS time of about 10-50 Myr for solar metallicity. 
We find that the
kinematics of this young population do not follow the expected signal from a long-lived stable warp.

Our approach is to compare the observations with the expectations derived from a model of the distribution and
kinematics of this young population of stars, taking into full account the known properties of the astrometric errors,
thereby avoiding the biases that can be introduced by using intrinsically uncertain and biased distances to
derive unobserved quantities. In Section 2 we describe the data for our selected sample of OB stars from
HIP2 and from \emph{Gaia} DR1. In Section 3 we present the model developed to create mock catalogues reproducing the observed
distributions. In Section 4 we report the results of comparing the proper motion distributions of our two samples with
the probability distribution of the proper motions derived from models with and without a warp. In the last sections we
discuss the possible implications of our results and outline future steps.


\section{The data \label{SecData}}

In this contribution we will analyse the both the pre-\emph{Gaia} astrometry from the New \emph{Hipparcos}
Reduction \citep{VanLeeuwen2007}, as well as from \emph{Gaia} DR1 \citep{Brown:2016}. First, our approach here to
analyzing the proper motions is significantly different from that used previously by \cite{Smart:1998} and
\cite{Drimmel:2000} for the first \emph{Hipparcos} release; any new results based on new \emph{Gaia} data cannot be
simply attributed to better data or better methods. Also a study of the \emph{Hipparcos} error properties is necessary
for understanding the astrometric error properties of the \emph{Hipparcos} subsample in \emph{Gaia} DR1  that we
consider here because of the intrinsic connection, by construction, between the astrometry of the New \emph{Hipparcos}
Reduction and \emph{Gaia} DR1 \citep{Michalik:2015}. Finally, the two samples are complementary, as the
\emph{Hipparcos} sample can be considered more complete and is substantially larger than the \emph{Hipparcos} subsample
of OB stars in \emph{Gaia} DR1 with superior astrometry, as explained below.

We select from the New \emph{Hipparcos} Catalogue (hereafter HIP2) the young OB stars, due to their high intrinsic luminosity.
Moreover, being short-lived, they are expected to trace the warped gaseous component. However, the spectral types in
the HIP2 are simply those originally provided in the first \emph{Hipparcos} release. In the hope that the many stars
originally lacking luminosity class in the \emph{Hipparcos} catalogue would have by now received better and more complete
spectral classifications, we surveyed the literature of spectral classifications available since the \emph{Hipparcos} release.
Most noteworthy for our purposes is the Galactic O-star Spectroscopic Survey (GOSSS) \citep{MaizApellaniz:2011,
Sota:2011, Sota:2014}, an ongoing project whose aim is to derive accurate and self-consistent spectral types of all
Galactic stars ever classified as O type with $B_J$ magnitude $<12$. From the catalogue presented in \cite{Sota:2014},
which is complete to $B_J=8$ but includes many dimmer stars, we imported the spectral classifications for the 212 stars
that are present in the HIP2 catalogue. Thirteen of these HIP2 sources were matched to multiple GOSSS sources, from
which we took the spectral classification of the principle component. Also worth noting is the Michigan Catalogue of HD
stars \citep{MichCat1,MichCat2,MichCat3,MichCat4,MichCat5}, which with its 5th and most recent release now covers the
southern sky ($\delta < 5^\circ$), from which we found classifications for an additional 3585 OB stars. However, these
two catalogues together do not cover the whole sky, especially for the B stars. We therefore had to resort to tertiary
sources that are actually compilations of spectral classifications, namely the Catalogue of Stellar Spectral
Classifications (4934 stars; \cite{Skiff:2014}), and the Extended Hipparcos Compilation (3216 stars;
\cite{Anderson:2012}). In summary, we have spectral classifications for 11947 OB stars in \emph{Hipparcos}.

We select from the HIP2 only those stars with spectral type earlier than B3,
with an apparent magnitude $V_J \le 8.5$, 
and with galactic latitude $|b| < 30^o$, resulting in 1848
OB stars. From this sample of HIP2 stars we define two subsamples: a HIP2 sample whose measured \emph{Hipparcos} parallax is
less than 2 mas, and a TGAS(HIP2) sample consisting of those HIP2 stars that appear in the \emph{Gaia} DR1 whose measured TGAS
parallax is less than 2 mas. The cut in parallax, together with the cut in galactic latitude, is done to remove local
structures (such as the Gould Belt). Our HIP2 sample contains 1088 stars (including 18 stars without luminosity class),
while our TGAS(HIP2) sample contains only 788 stars. This lack of HIP2 in TGAS stars is largely due to the
completeness characteristics of DR1, discussed further in Section \ref{completeness} below.

\begin{figure}[ht]
  \resizebox{\hsize}{!}{\includegraphics{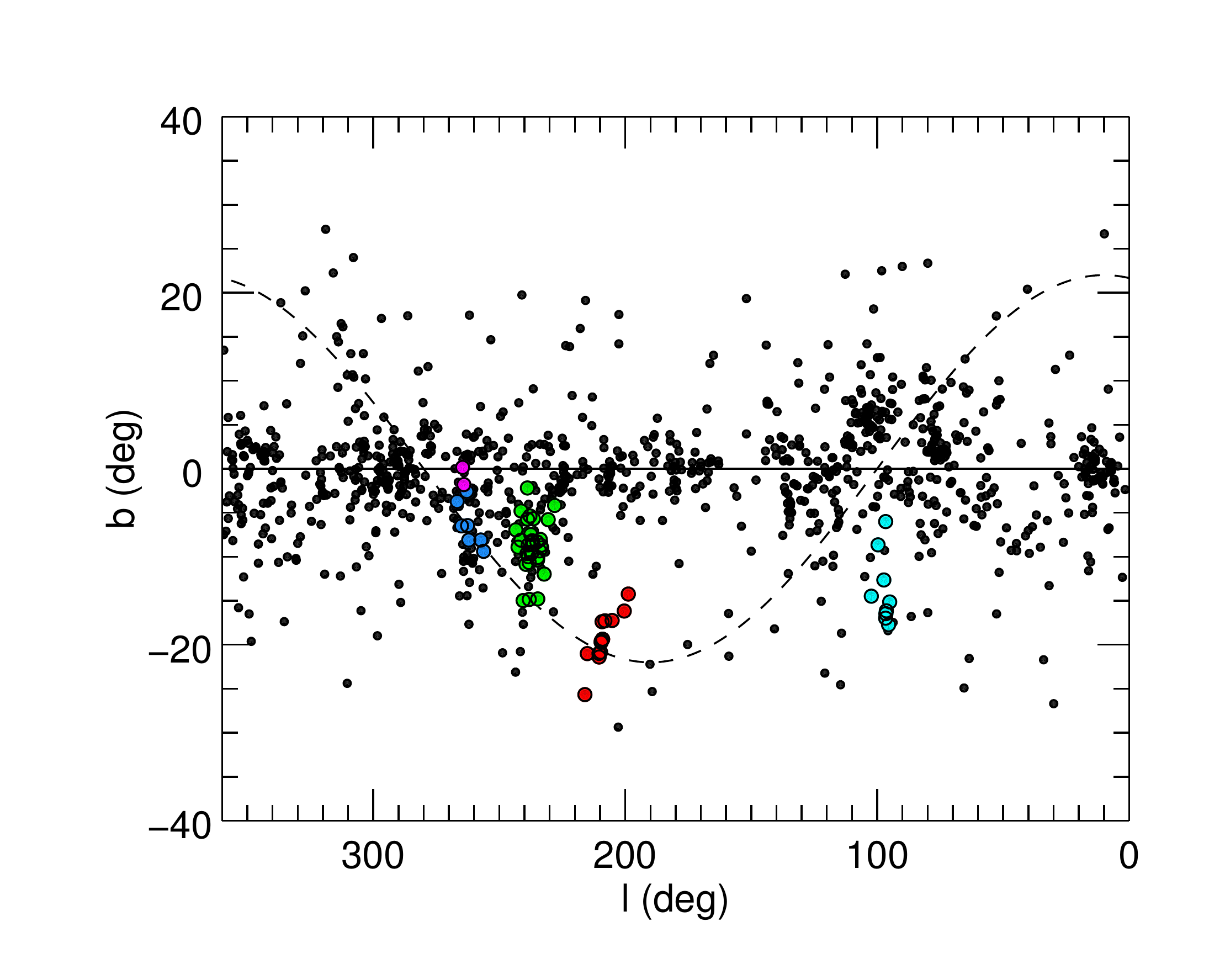}}
  \caption{Our final sample of \emph{Hipparcos} OB stars on the sky, plotted in galactic coordinates. The dashed line shows
  the orientation of the Gould belt according to \cite{Comeron:1992}. Colored points indicate the stars that are
  identified members of the OB associations Orion OB1 (red), Trumpler 10 (purple), Vela OB2 (blue), Collinder 121 (green) and Lacerta OB1 (cyan).
}
  \label{lbplot}
\end{figure}

Notwithstanding the parallax cut we found that there were HIP2 stars in our sample that are members of nearby OB
associations known to be associated with the Gould Belt. We therefore removed from the HIP2 sample members of the Orion
OB1 association (15 stars, as identified by \cite{Brown:1994}) and, as according to \cite{DeZeeuw:1999}, members of the
associations Trumpler 10 (2 stars), Vela OB2 (7 stars),  Lacerta OB1 (9 stars), all closer than 500 pc from the Sun. We
also removed 33 stars from the Collinder 121 association as it is thought to also be associated with the Gould Belt.
With these stars removed we are left with a final HIP2 sample of 1022 stars. It is worth noting that the superior \emph{Gaia}
parallaxes already result in a cleaner sample of distant OB stars: only 21 members of the above OB associations needed
to be removed from the TGAS(HIP2) sample after the parallax cut. Figure \ref{lbplot} shows the position of the stars in
our two samples in Galactic coordinates.

Due to the above mentioned parallax cut, our sample mostly contains stars more distant than 500 pc.
Though our analysis will only marginally depend on the distances, we use spectro-photometric parallaxes when a distance
is needed for the HIP2 stars, due to the large relative errors on the trigonometric parallaxes. Absolute
magnitudes and intrinsic colors are taken from \cite{Martins:2005aa} and \cite{Martins:2006} for the O stars, and from
\cite{Humphreys:1984} and \cite{Flower:1977} for the B stars.

We note that the OB stars with $V_J \le 7.5$ (approximately 90\% complete) beyond the Solar Circle ($90 \degr < l
< 270 \degr$) show a tilt with respect to the Galactic plane that is consistent with a Galactic warp: a robust linear
fit in the l-b space (l normalized to $180 \degr - l$) yields a slope of $0.049 \pm 0.007$. However, given the possible
effects of patchy extinction, it would be risky to make any detailed conclusions about the large scale geometry of
the warp from this sample with heliocentric distances limited to a few kiloparsecs.


\section{The model \label{secmodel}}

Here we describe the model used to produce synthetic catalogues and probability distribution functions of the observed
quantities to compare with our two samples of \emph{Hipparcos} OB-stars described above, taking into account the error
properties of the \emph{Hipparcos} astrometry (for the HIP2 sample) and of the Hipparcos subsample in Gaia DR1 (for the
TGAS(HIP2) sample), and applying the same selection criteria used to arrive at our two samples.  The distribution on
the sky and the magnitude distribution of the HIP2 sample to $V=7.5$ (556 stars), assumed to be complete, are
first reproduced in Sections \ref{SecLumFunc} and \ref{SecSpatDistr}, using models of the color-magnitude and spatial
distribution of the stars, and a 3D extinction model. The completeness of the \emph{Hipparcos}, and of TGAS with respect
to \emph{Hipparcos}, is described in Section \ref{completeness} and is used to model our two samples down to $V=8.5$.
Then a simple kinematic model for the OB
stars is used to reproduce the observed distribution of proper motions (Section \ref{SecKin}) of our two samples,
including (or not) the expected effects of a stable (long-lived) non-precessing warp (Section \ref{SecWarp}). A
comparison of the observed samples with the expectations from the different warp/no-warp models is presented in Section
\ref{results}.

The model that we present here is purely empirical. Many parameters are taken from the literature, while a limited
number have been manually tuned when it was clear that better agreement with the observations could be reached. We
therefore make no claim that our set of parameters are an optimal set, nor can we quote meaningful uncertainties. The
reader should thus interpret our choice of parameters as an initial "first guess" for a true parameter adjustment,
which we leave for the future when a larger dataset from \emph{Gaia} is considered. In any case, after some exploration, we
believe that our model captures the most relevant features of the OB stellar distribution and kinematics at scales
between 0.5 -- 3 kpc.

\subsection{Warp \label{SecWarp}}

In this Section we describe our model for the warp in the stellar spatial distribution and the associated kinematics.
In Section \ref{SecSpatDistr} below we construct a flat-disk distribution with vertical exponential profile, then
displace the z-coordinates by $z_w$, where:
\begin {equation}
\label{zwarp} z_w(R,\phi) = h(R) \, \sin{(\phi + \phi_w)} \quad.
\end{equation}
The warp phase angle $\phi_w$ determines the position of the line-of-nodes of the warp with respect to the
galactic meridian ($\phi = 0$). The increase of the warp amplitude with Galactocentric radius, $h(R)$, is
described by the height function
\begin {equation}
\label{hzwarp} h(R)=h_0 \, (R-R_w)^{\alpha_w} \quad,
\end{equation}
where $h_0$ and $R_w$ are the warp amplitude and the radius at which the Galactic warp starts, respectively. The
exponent $\alpha_w$ determines the warp amplitude increase. Table \ref{tab:warp_models} reports three different sets of
warp parameters taken from the literature and used later in our analysis in \ref{Seclmub}.
\begin{table*}[ht]
   \caption{
   Comparison of warp parameters for the models of \cite{Drimmel:2001} and \cite{Yusifov:2004}.
   The radius $R_w$ was scaled to account for different assumptions about the Sun - Galactic center distance
   in this work and in the considered papers.
     }
  \label{tab:warp_models}
  \centering
  \begin{tabular}{l c c c c}
    \hline
    \hline
             & $R_w$ (kpc) & $\alpha_w$ & $h_0$(kpc$^{\alpha_w -1}$) & $\phi_w(\degr)$\\
    \hline
     \cite{Drimmel:2001}, dust             &  7 & 2 & 0.073 & 0\\
     \cite{Drimmel:2001}, stars            & 7 & 2 & 0.027 & 0\\
     \cite{Yusifov:2004}                      & 6.27 & 1.4 & 0.037 & 14.5\\
     \hline
  \end{tabular}
\end{table*}
Assuming that the Galaxy can be
modelled as a collisionless system, the $0^{th}$ moment of the collisionless Boltzmann equation in cylindrical
coordinates gives us the mean vertical velocity $\bar{v}_z (R,\phi)$. If we use the warped disk as described above and
suppose that $\bar{v}_R = 0 $ (i.e. that the disc is not radially expanding or collapsing), we obtain:
\begin {equation}
\label{vz_general} \bar{v}_z (R,\phi)=\frac{\bar{v}_\phi}{R}  \, h(R) \, \cos{( \phi + \phi_w )} \quad.
\end{equation}
Equations \ref{zwarp}, \ref{hzwarp} and \ref{vz_general} assume a perfectly static warp.  It is of course possible to
construct a more general model by introducing time dependencies in Equations \ref{zwarp} and \ref{hzwarp}, which will
result in additional terms in Equation \ref{vz_general}, including precession or even an oscillating (i.e. "flapping")
amplitude. For our purpose here, to predict the expected systematic vertical velocities associated with a warp, such
time dependencies are not considered.

This above model we refer to as the \emph{warp model}, with three possible sets of parameters reported in Table
\ref{tab:warp_models}. Our alternative model with $z_w = 0$ will be the \emph{no-warp model}, where Equation
\ref{vz_general} reduces to the trivial $\bar{v}_z=0$.

\begin{figure}[ht]
\resizebox{\hsize}{!}{\includegraphics{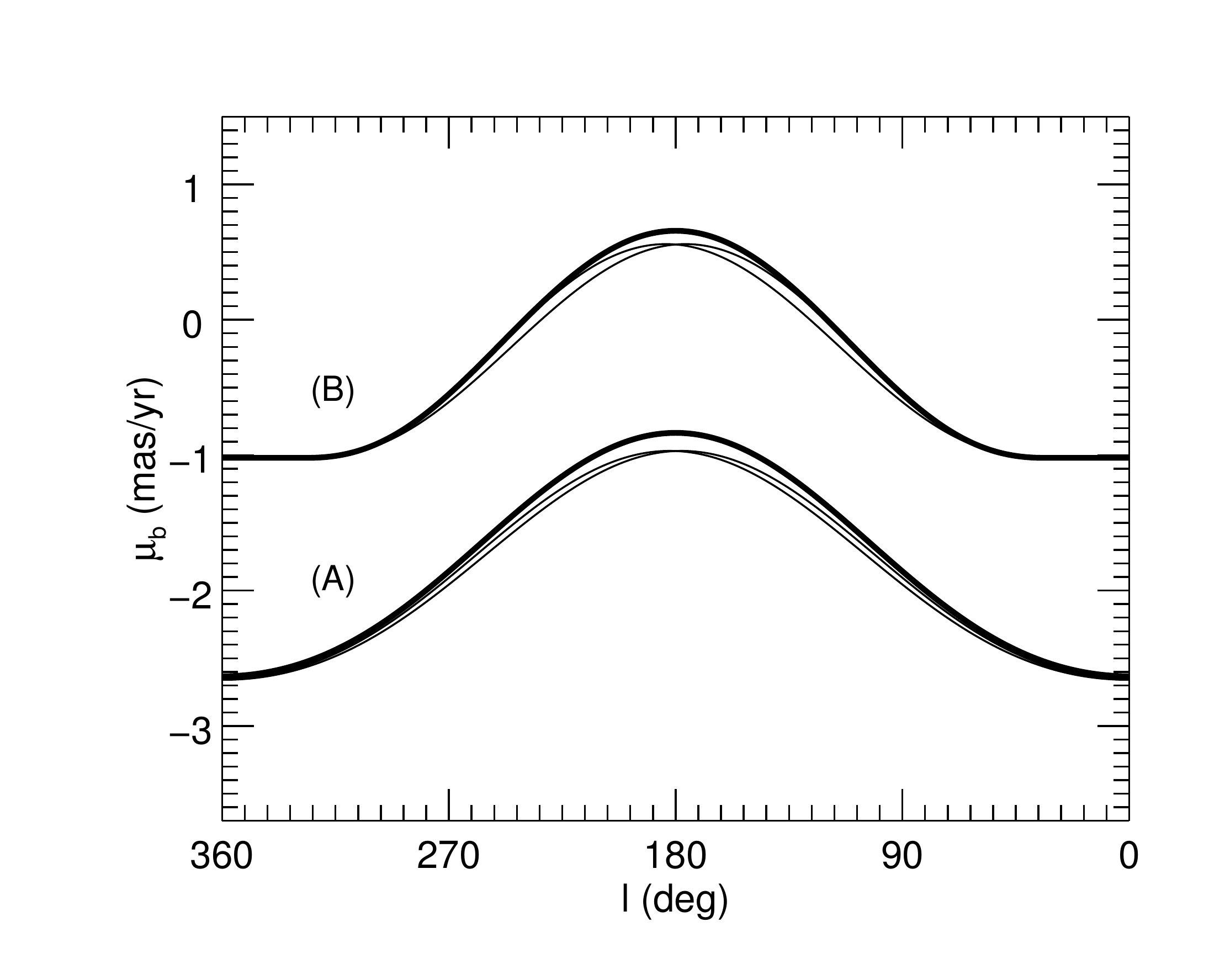}}
\caption{ According to the warp model, the true $\mu_b$ in the Galactic plane as a function
of Galactic longitude at heliocentric distances of 0.5 kpc (A) and 1.5 kpc (B). For each set of curves,
the thick line represents the case with warp phase $\phi_w=0 \degr$ and the two thin curves
show $\phi_w= \pm 20 \degr$.
}
\label{mubth}
\end{figure}

Figure \ref{mubth} shows the prediction of a warp model, without errors, for the mean proper motions $\mu_b$ in
the Galactic plane. For the no-warp model (here not shown), we expect to have negative $\mu_b$ values symmetrically
around the Sun as the reflex of the vertical component of Solar motion, progressively approaching 0 with increasing
heliocentric distance. For a warp model, a variation of $\mu_b$ with respect to galactic longitude is introduced,
with a peak toward the anti-center direction ($l=180 \degr$). Figure \ref{mubth}
also shows that a variation of the warp phase angle has a rather minor effect on the kinematic signature, nearly
indistinguishable from a change in the warp amplitude.

\subsection{Luminosity function \label{SecLumFunc}}

There are different initial luminosity functions (ILF) in the literature for the upper main sequence
\citep{BahcallSoneira:1980, Humphreys:1984, Scalo:1986, Bahcall:1987, Reed:2001}. Given the uncertainties in the ILF
for intrinsically bright stars (absolute magnitude $M < -3$), we assume $N(M) \propto 10^{\, \alpha M}$, and use the
value $\alpha=0.72$ that we find reproduces well the apparent magnitude distribution (Figure \ref{MagDistr}) with the
spatial distribution described in Section \ref{SecSpatDistr}. We use a main sequence Color-Magnitude relation
consistent with the adopted photometric calibrations (see Section \ref{SecData}). Absolute magnitudes $M$ are randomly
generated consistent with this ILF then, for a given absolute magnitude, stars are given an intrinsic color generated
uniformly inside a specified width about the main sequence (Figure \ref{ColMagFig}), which linearly increases as stars
get fainter. According to an assumed giant fraction (see below), a fraction of the stars are randomly labelled as
giant. The absolute magnitude of these stars are incremented by $-0.5$ mag, and their color is generated uniformly
between the initial main sequence color and the reddest value predicted by our calibrations, $(B-V)_0=-0.12$. The giant
fraction $f_g$ has been modelled as a function of the absolute magnitude as follows:
$$
f_g(M)=
\begin{cases}
    1,& \text{if } M\leq -7\\
    -0.25 M - 0.75,& \text{if } -7 \leq M\leq -4 \,\\
    0.25,              & \text{if } M\geq -4
\end{cases} \quad,
$$
in order to roughly reproduce the fraction of giants in the observed catalogue. We caution that this procedure is not
intended to mimic stellar evolution. Instead, we simply aim to mimic the intrinsic color-magnitude distribution (i.e.
Hess diagram) of our sample.

\begin{figure}[ht]
\resizebox{\hsize}{!}{\includegraphics{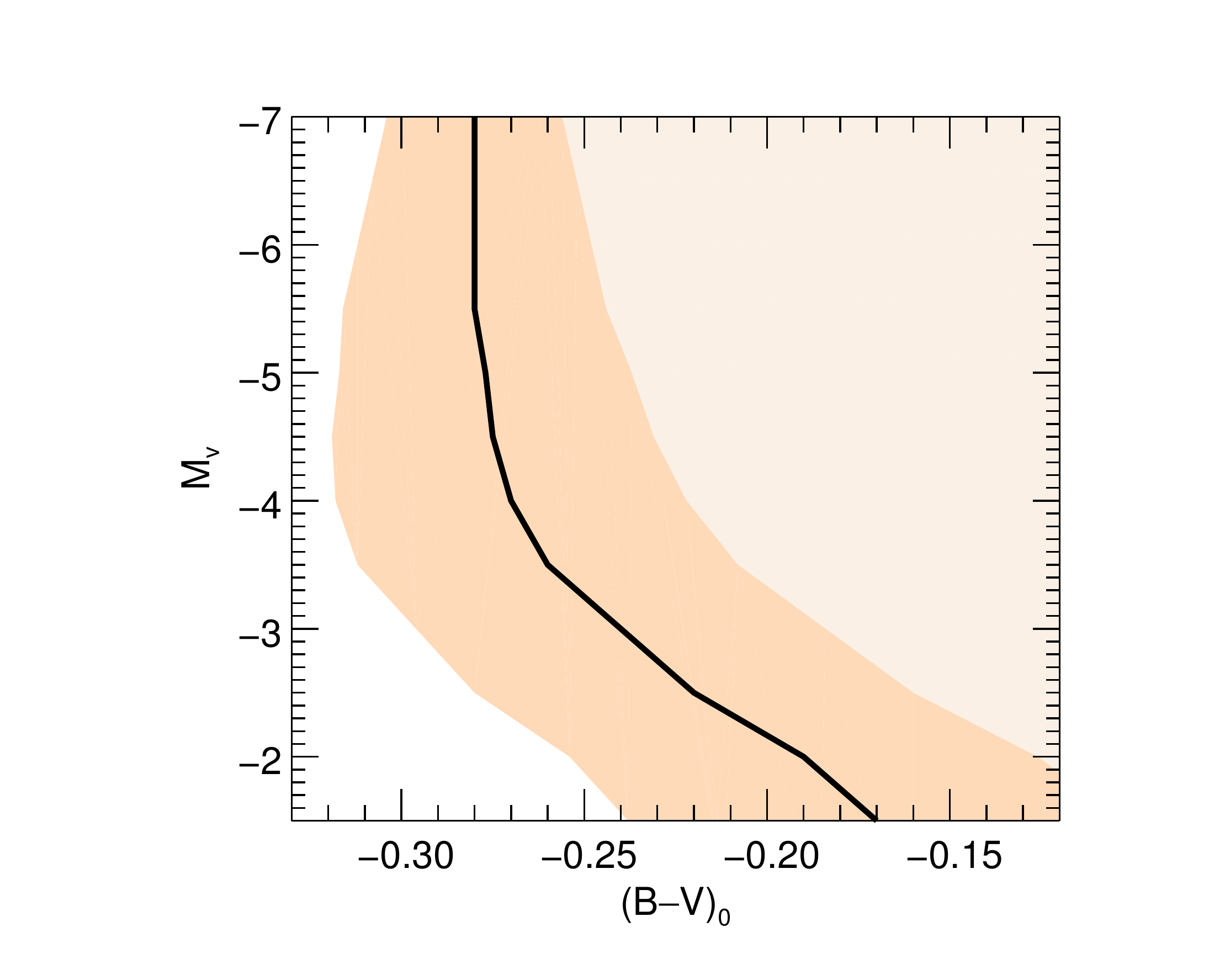}} 
\caption{Color-Magnitude diagram used to generate synthetic
intrinsic colors. The dark and light orange regions shows, respectively, the main sequence and the giant regions. The
density of the two regions (here not shown) depends on the ILF and on the giant fraction. } \label{ColMagFig}
\end{figure}

\subsection{Spatial distribution and extinction}
\label{SecSpatDistr}

\begin{figure}[ht]
  \resizebox{\hsize}{!}{\includegraphics{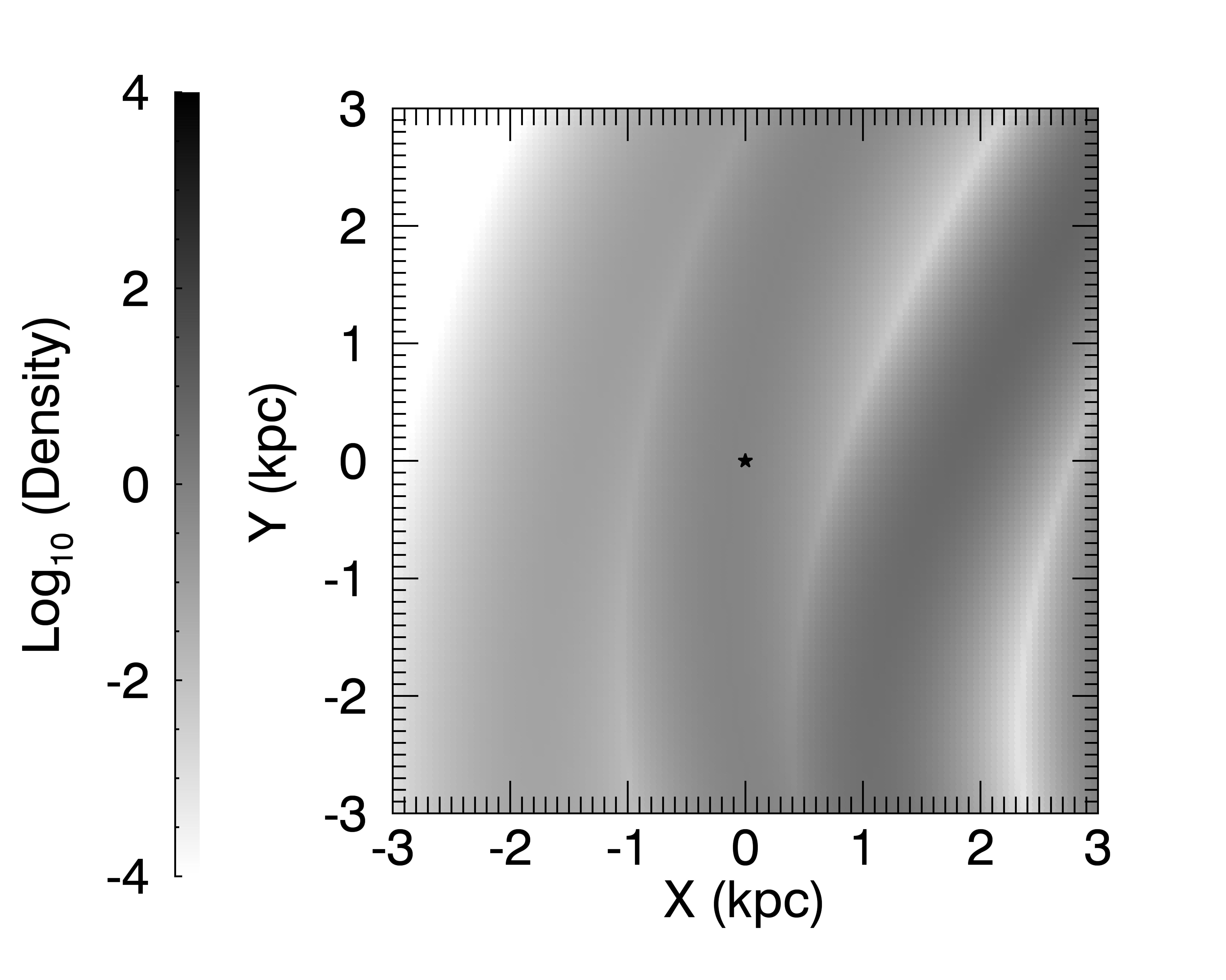}}
  \caption{Modelled Surface density of the OB stars. Sun's position is indicated by the star.}
  \label{DensSpArms}
\end{figure}

Since we wish to model the distribution of OB stars on scales larger than several hundred parsecs, we use a
mathematical description of this distribution that smooths over the inherent clumpy nature of star formation, which is
evident if we consider the distribution of young stars within 500 pc of the Sun. On these larger scales it is
nevertheless evident that the OB stars are far from being distributed as a smooth exponential disk, but rather trace
out the spiral arms of the Galaxy, being still too young to have wondered far from their birth-places. In our model we
adopt $R_0=8.2$ kpc as the Sun's distance from the Galactic center, and a solar offset from the disk midplane of $z_0 =
25$ pc, for galactocentric cylindrical coordinates $(R, \phi, z)$, as recommended by \cite{BlandGerhard:2016}. For the
spiral arm geometry we adopt the model of \cite{Georgelin:1976}, as implemented by \cite{Taylor:1993},
rescaled to $R_0=8.2$ kpc, with the addition of a
local arm described as a logarithmic spiral segment whose location is described by $R_{Loc} = R_{Loc, 0} \exp{ -(
\tan{p} \phi)}$, $p$ being the arm's pitch angle. The surface density profile across an arm is taken to be gaussian,
namely: $\rho \propto \exp{(-d^2_a/w^2_a)}$ , where $d_a$ is the distance to the nearest arm in the $R,\phi$ plane, and
$w_a=c_w \, R$ is the arm half-width, with $c_w = 0.06$ \citep{Drimmel:2001}. An "overview" of the modelled surface
density distribution is shown in Figure \ref{DensSpArms}.
The stars are also given an exponential vertical scale height $\rho \propto \exp{(- |z'|/h_z)}$, where $h_z$ is the
vertical scale height and $z' = z - z_w - z_{Loc}$, $z_w(R,\phi)$ being the height of the warp as described in section
\ref{SecWarp}, and $z_{Loc}$ is a vertical offset applied only to the local arm.

We generate the above spatial distribution in an iterative Monte-Carlo fashion. Ten thousand positions in $(x,y)$
coordinates are first generated with a uniform surface density to a limiting heliocentric distance of 11 kpc, and with
an exponential vertical profile in $|z'|$. The relative surface density $\Sigma(x,y)$ is evaluated at each position according to our model described above, and
positions are retained if $u < \Sigma(x,y)/max(\Sigma)$, where $u$ is a uniform random deviate between 0 and 1. Each
retained position is assigned to a $(M_V,(B-V)_0)$ pair, generated as described in the previous section. The extinction
to each position is then calculated using the extinction map from \cite{Drimmel:2003} and the apparent magnitude is derived.
Based on the apparent magnitude, a fraction of the stars with $V \le 8.5$ are randomly retained consistent with
the completeness model of \emph{Hipparcos}, while for a TGAS-like catalogue an additional random selection of stars is
similarly made as a function of the observed apparent magnitude and color, as described in the following section.
This procedure is iterated until a simulated catalogue of
stars is generated matching the number in our observed sample.

\begin{figure}[ht]
\resizebox{\hsize}{!}{\includegraphics{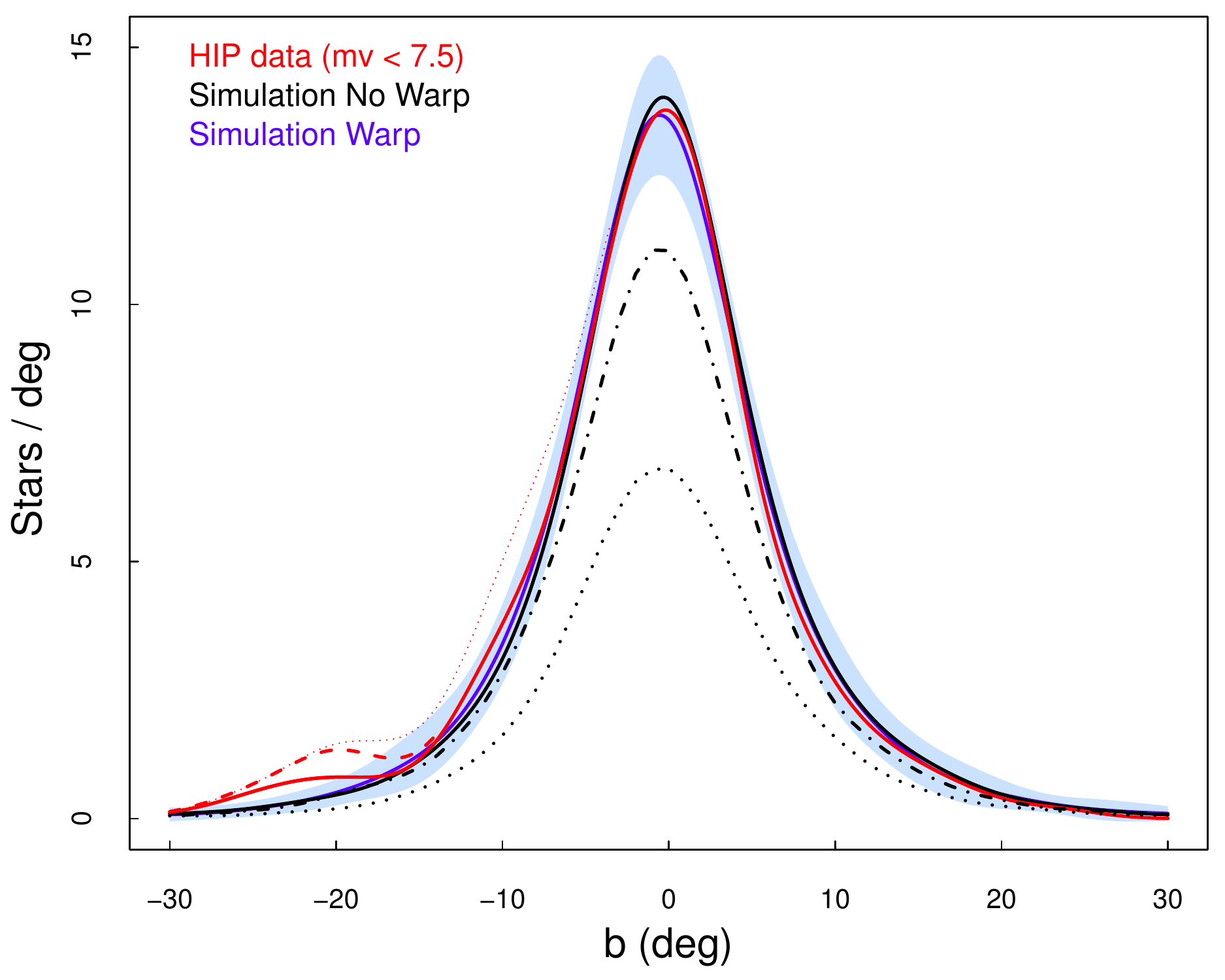}} 
\caption{Latitude distribution of the HIP2 OB stars. The red
curve is a non-parametric fit to the selected HIP2 sample, the red dashed curve shows the additional contribution of
the Orion OB1 association, while the red dotted the added contributions of the Trumpler 10, Vela OB2, Collinder 121 and
Lacerta OB1 associations. The blue curve and light-blue shaded area shows the average and 2 $\sigma$ confidence band of
the simulated longitude distribution, based on 30 simulated instances of the sample. The black dotted and dash-dot
curves show the relative contributions of the major spiral arms (Sagittarius-Carina and Perseus) and the local arm,
respectively, while the additional black solid curve is for a model without a warp. } \label{LatDistr}
\end{figure}

\begin{figure}[ht]
\resizebox{\hsize}{!}{\includegraphics{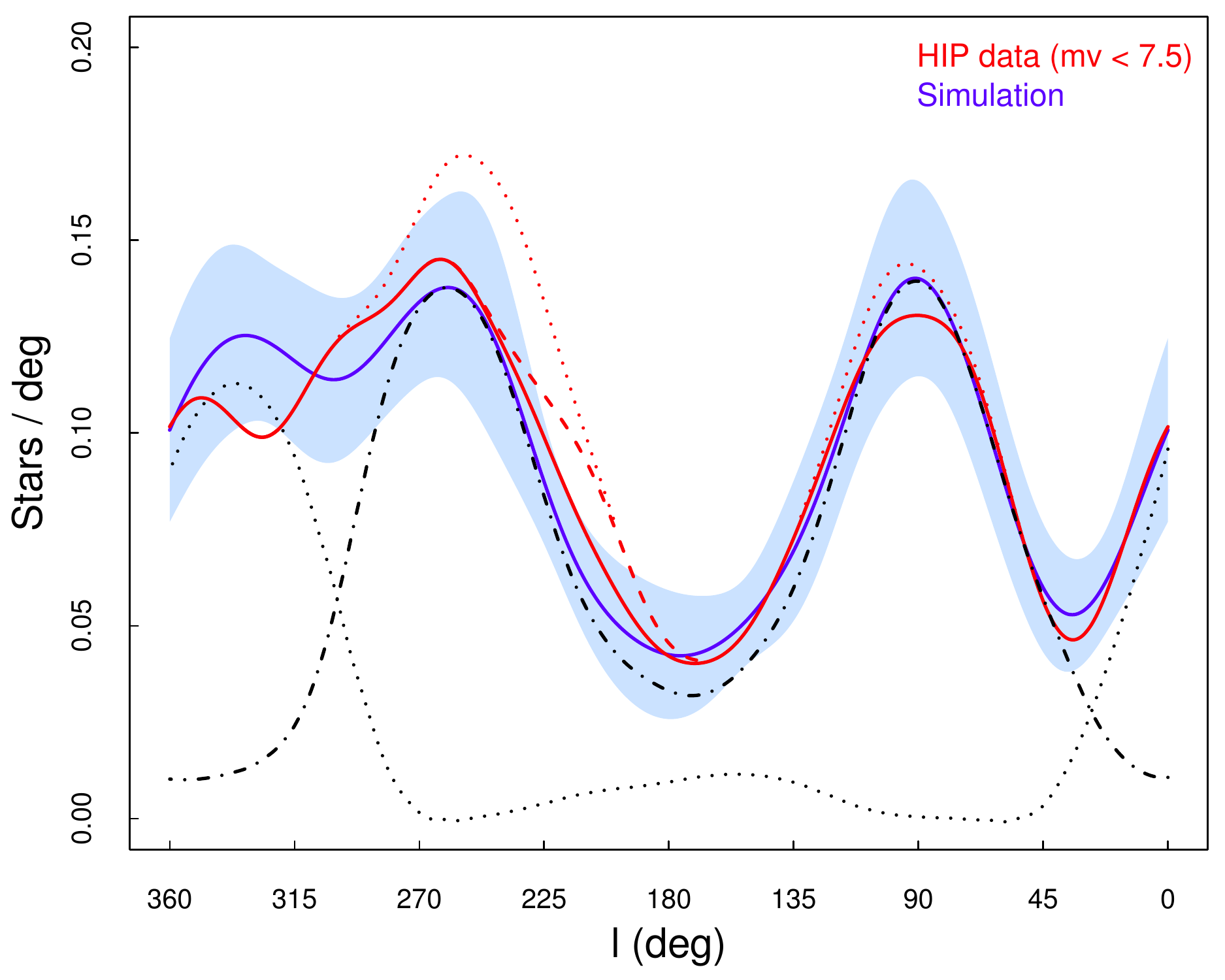}} 
\caption{Longitude distribution of the HIP2 OB stars. Meaning
of the curves are as in the previous figure \label{LongDistr}.}
\end{figure}

Good agreement with the HIP2 distribution in galactic latitude was found adopting a vertical scale height $h_z=70$ pc
and assuming $z_{Loc} = 25 \, \text{pc}$, as shown in Figure \ref{LatDistr}. Figure \ref{LongDistr} compares the
modelled and observed distributions in galactic longitude, which is dominated by the local arm.  This observed
distribution is reproduced by placing the local arm at a radius of $R_{Loc,0}=8.3$ kpc, with a pitch angle of
$6.5^{\circ}$ and a half-width of $500$ pc. (The curves in Figures \ref{LatDistr} and \ref{LongDistr} are
non-parametric fits to the distributions obtained through kernel density estimation with a gaussian kernel, as
implemented by the generic function \emph{density} in R \footnote{https://www.r-project.org}. The smoothing bandwith is
fixed for all the curves in the same figure, with values of $2.5^{\circ}$ and $15^{\circ}$ for the latitude and the
longitude distribution, respectively.)


\begin{figure}[ht]
\resizebox{\hsize}{!}{\includegraphics{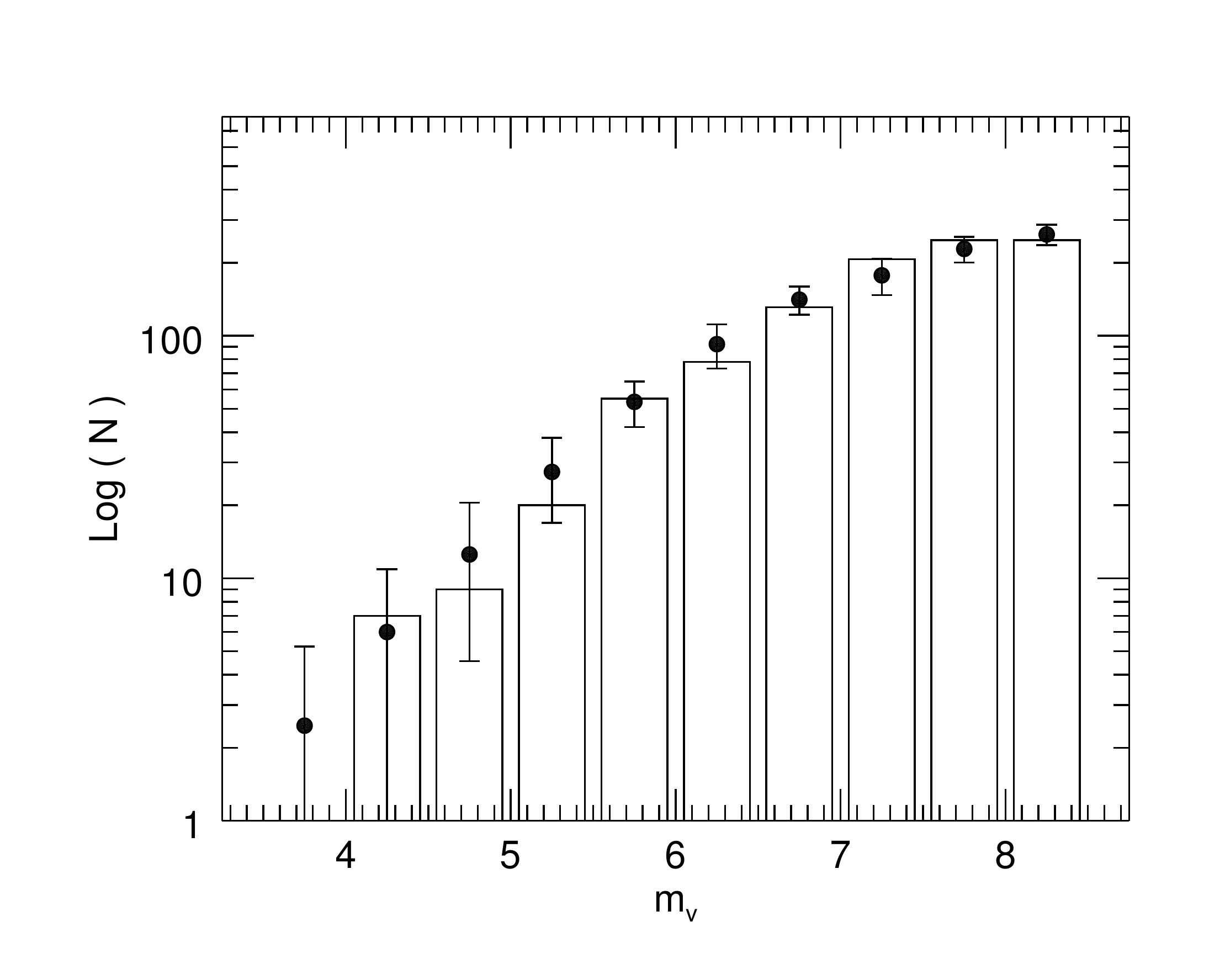}} 
\caption{Apparent magnitude distribution for the data
(histogram) and the simulations (black dots). The error bars show 2 $\sigma$ uncertainty, calculated with 30 simulated
samples.} \label{MagDistr}
\end{figure}

Figure \ref{MagDistr} shows the resulting apparent magnitude distribution, as compared to the HIP2 sample.
Comparing the observed and the simulated longitude distributions in Figure \ref{LongDistr}, we note that our model
fails to reproduce well the observed distribution in the longitude range $l=300-360$ degrees. This is probably
revealing a deficit in the geometry adopted for the Sagittarius-Carina arm, which we have not attempted to modify as we
are primarily interested in the kinematics toward the Galactic anti-center. It should also be noted that, for both the
longitude and latitude distributions, the presence or absence of a warp (modelled as described in Section
\ref{SecWarp}) has very little effect.

The careful reader will note that our approach assumes that the Hess diagram is independent of position in the Galaxy.
We recognize this as a deficit in our model, as the spiral arms are in fact star formation fronts, in general moving
with respect to galactic rotation. We thus expect offsets between younger and older populations, meaning that the Hess
diagram will be position dependent. However, if the Sun is close to co-rotation, as expected, such offsets are minimal.

\subsection{Completeness \label{completeness}}

\begin{figure}[ht]
\resizebox{\hsize}{!}{\includegraphics{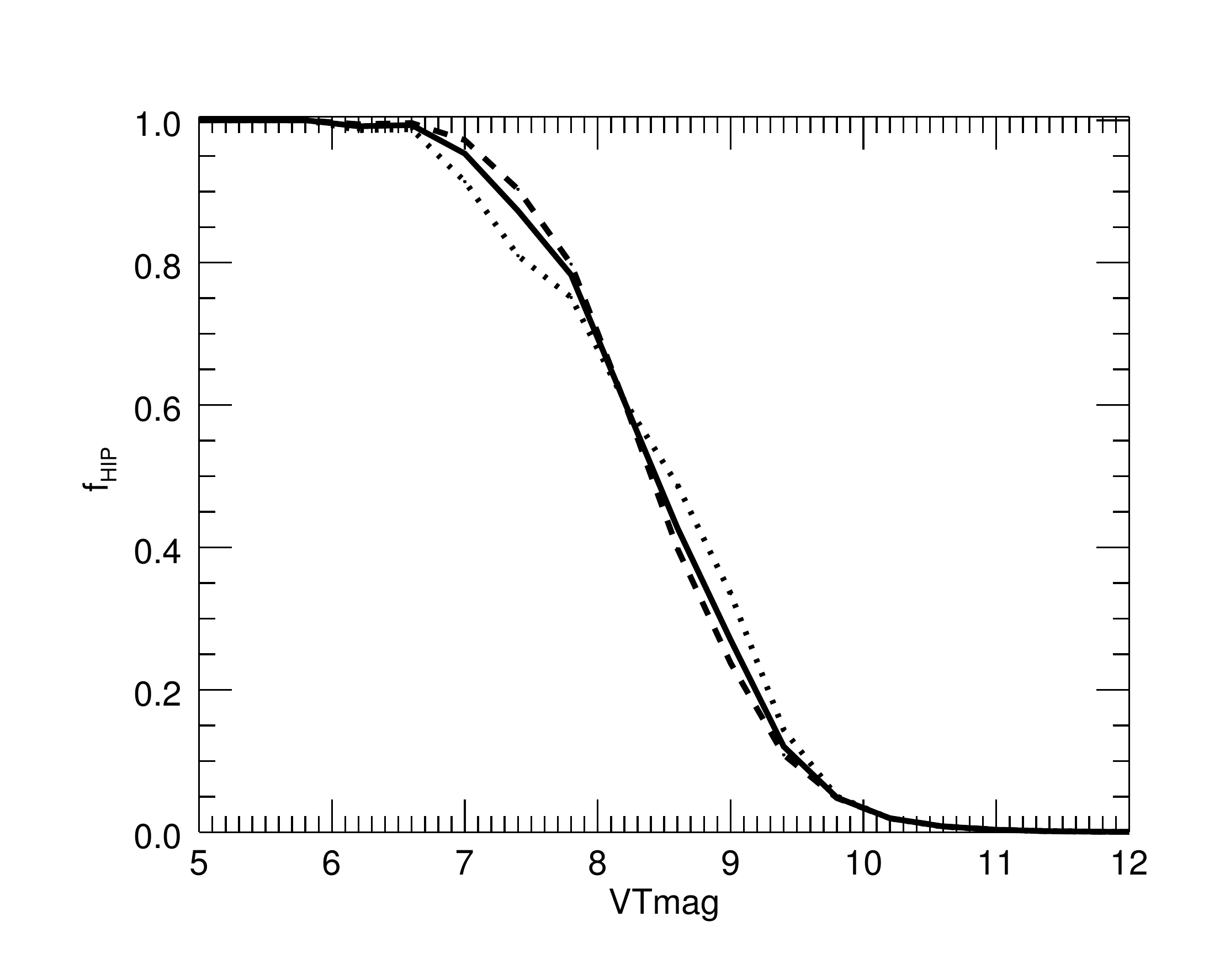}}
\caption{Fraction of \emph{Hipparcos} completeness in function of apparent magnitude $V_T$ with respect to the Tycho-2 catalogue.
The dashed and the dotted line represent, respectively, the \emph{Hipparcos} fraction for the stars above and below $\delta = -30^o$.
}
\label{compl_frac}
\end{figure}

The completeness of the \emph{Hipparcos} catalogue decreases with apparent magnitude, as shown by the fraction of
Tycho-2 stars in the \emph{Hipparcos} catalogue (Figure \ref{compl_frac}). At $V_T \leq 7.5$ the HIP2 catalogue is
approximately 90\% complete, reaching approximately 50\% completeness at $V_T = 8.5$.  The fact that the
\emph{Hipparcos} catalogue was based on an input catalogue built from then extant ground-based surveys results in
inhomogenous sky coverage. In particular we find a north/south dichotomy at Declination $\delta \approx -30\degr$.  We
assume that the completeness fraction of \emph{Hipparcos} stars in function of $V_T$ decreases in a similar way for the
Johnson magnitude, i.e. $f_{HIP} (V_T) \approx f_{HIP} (V_J)$.

\begin{figure}[ht]
\resizebox{\hsize}{!}{\includegraphics{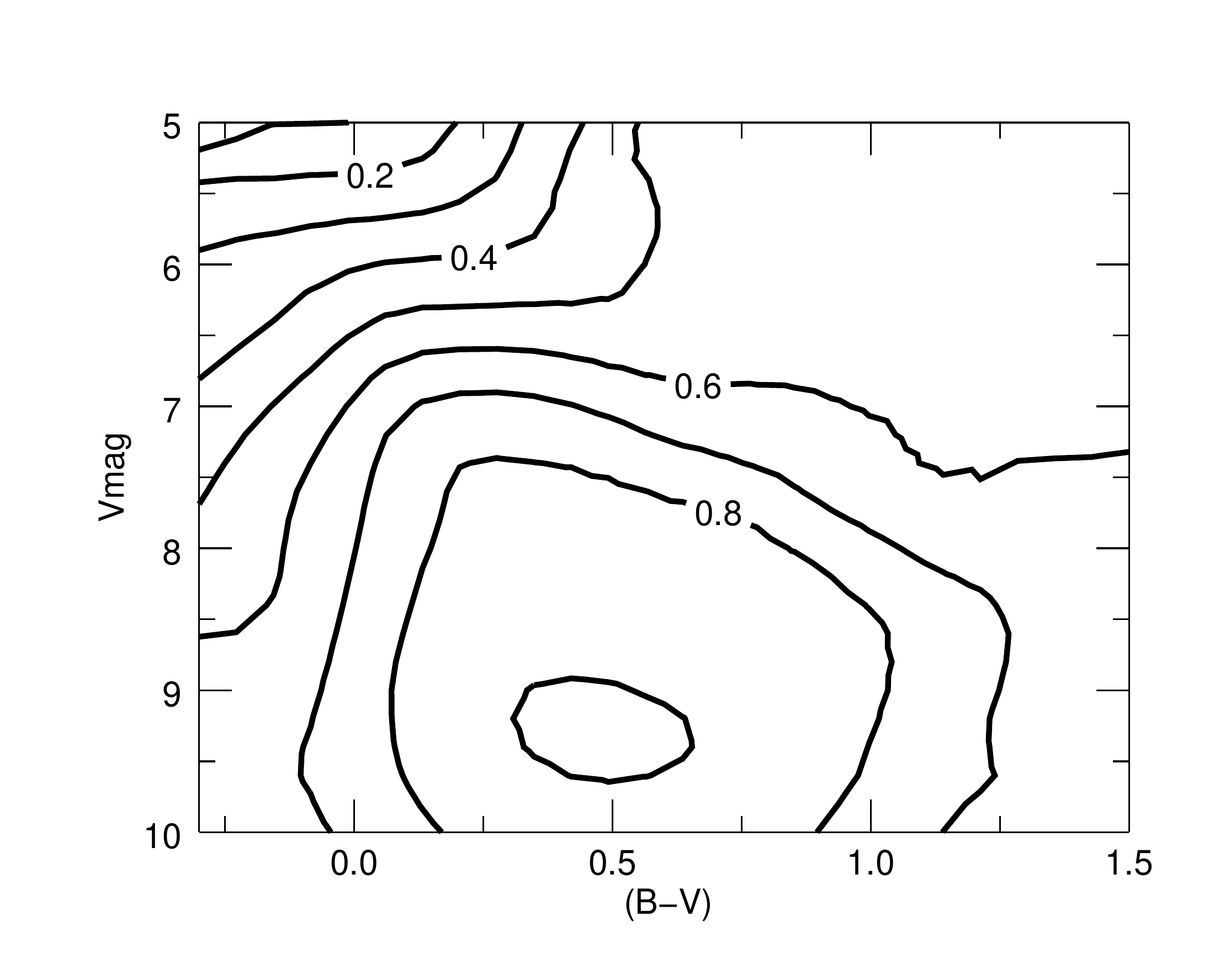}} 
\caption{ Fraction of HIP2 OB stars present in
HIP-TGAS as a function of the observed color and the apparent magnitude. } \label{compl_map}
\end{figure}

As already noted in Section 2, TGAS contains only a fraction of the stars in \emph{Hipparcos}. We find that the
completeness of TGAS with respect to \emph{Hipparcos} is strongly dependent on
the observed magnitude and color of the stars: 50\% completeness is reached at $V_J = 6.5$ mag and $B-V = 0$, with the
brightest and bluest stars missing from TGAS. This incompleteness is a result of the quality criteria
used for constructing TGAS and of the difficulty of
calibrating these stars due to their relative paucity. Figure \ref{compl_map} shows a map of the TGAS(HIP2)
completeness as a function of apparent magnitude and color. The completeness reaches a maximum plateau of about 80\%,
however this is not uniform across the sky. Due to the scanning strategy of the \emph{Gaia} satellite, and limited number of
months of observations that have contributed to the DR1, some parts of the sky are better covered than others. This
results in a patchy coverage, which we have not yet taken into account. However, this random sampling caused by the
incomplete scanning of the sky by \emph{Gaia} is completely independent of the stellar properties, so that our TGAS(HIP2)
sample should trace the kinematics of the stars in an unbiased way.

\begin{figure}[ht]
\resizebox{\hsize}{!}{\includegraphics{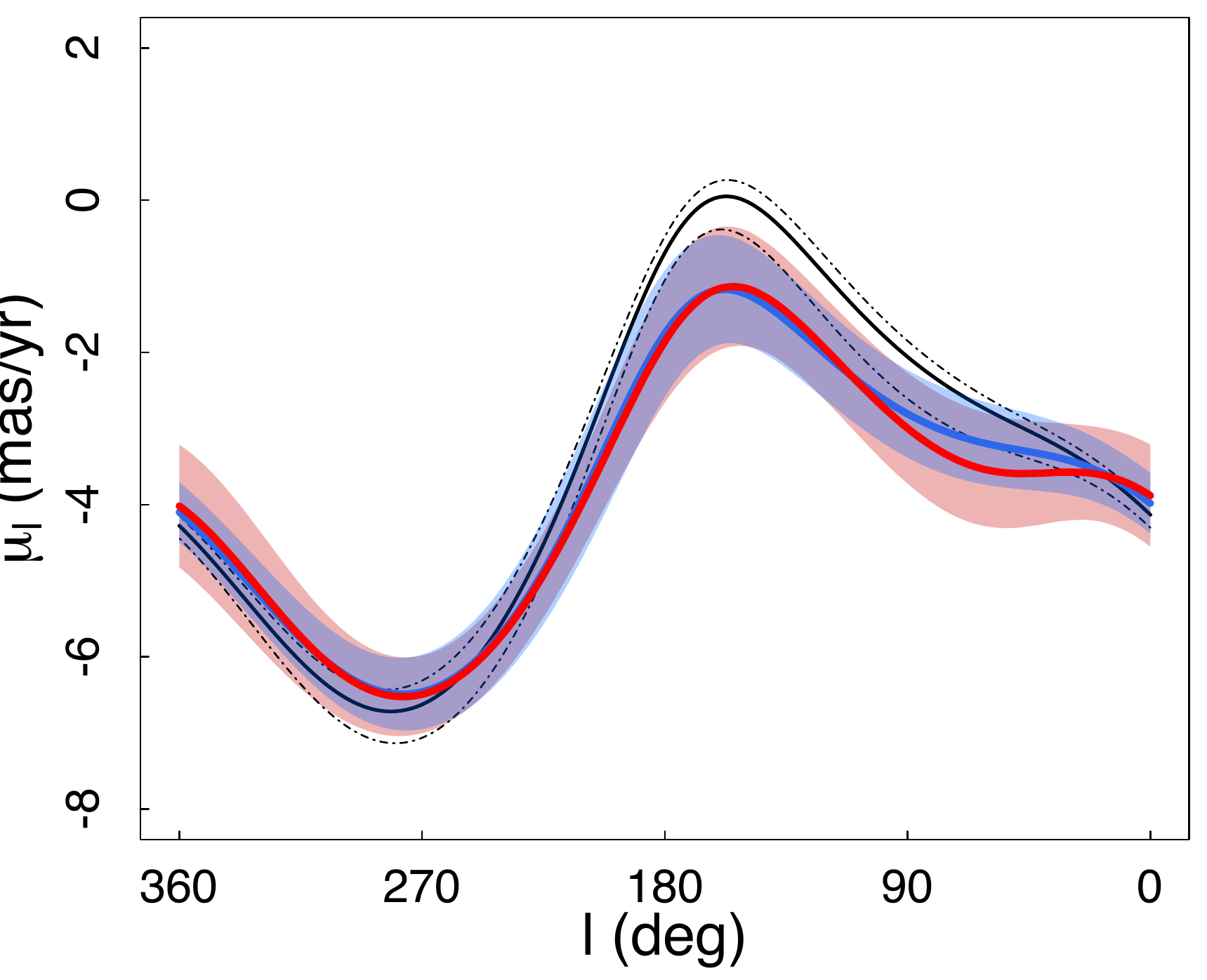}}
\caption{$\mu_l$ in function of Galactic
longitude for the data (red curve) together with the 95\% bootstrap confidence band (pink shaded area). The three black
dotted curved show the trend obtained with simulations with circular velocity 260, 238 and 220 km/s, respectively, from
the lowest to the highest curve. Simulations with an additional velocity to the Local Arm (see text) produce the
blue curve, for which the light blue shaded area shows the 2 $\sigma$ confidence band, calculated with 30 simulated catalogues.}
\label{Figlmul}
\end{figure}

\subsection{Kinematics}
\label{SecKin}
Now that the spatial distribution has been satisfactorily modelled, we can address the observed distribution of proper
motions. We point out that the kinematic model described in this Section is independent from the inclusion (or not)
of the warp-induced offset in latitude proper motions (Section \ref{SecWarp}).
We adopt a simple model for the velocity dispersions along the three
main axes of the velocity ellipsoid: $\sigma_{(1,2,3)}=\sigma^0_{(1,2,3)}  \, exp{ \Bigl( (R_{\odot}-R)/2  h_R  \Bigr)} $,
where $h_R=2.3$ kpc is the radial scale length and $\sigma^0_{(1,2,3)}=(14.35,9.33,5.45) \,\text{km s}^{-1}$
are the three velocity dispersions in the solar neighborhood
for the bluest stars \citep{Dehnen:1998}. A vertex deviation of $l_v=30^o$ is implemented, as measured by
\cite{Dehnen:1998} for the bluest stars, although we find that it has no significant impact on the proper motion
trends.

As recommended by \cite{BlandGerhard:2016}, we adopted $\Theta_0 = 238\, \text{km s}^{-1}$ for the circular rotation
velocity at the Solar radius $R_0$. Given that current estimates of the local slope of the rotation curve varies from
positive to negative values, and that our data is restricted to heliocentric distances of a few kpc, we assume a flat
rotation curve. In any case, we have verified that assuming a modest slope of $\pm 5$ km/s/kpc does not significantly
impact the expected trend in proper motions.

After local stellar velocities $(U,V,W)$ of the stars are generated, proper motions are calculated assuming a Solar
velocity of ${\bf v}_\odot = (U_{\odot},V_{\odot},W_{\odot})=(11.1,12.24,7.25) \,\text{km s}^{-1} $
\citep{Schoenrich:2010}. Observed proper motions in $(\alpha, \delta)$ are derived by adding random errors as per an
astrometric error model, described in the Section \ref{SecErr}. Finally, the proper motions in equatorial coordinates
are converted to galactic coordinates, ie. ($\mu_l*, \mu_b$), where $\mu_l* = \mu_l \cos b$.

Figure \ref{Figlmul} shows the derived proper motions in galactic longitude for both the data and simulations using the
bivariate local-constant (i.e. Nadaraya-Watson) kernel regression implemented by the \emph{npregbw} routine in the
\emph{np} R package with bandwidth $h=45^o$. The solid black line shows the trend obtained for the simulation with the
above listed standard parameters. Our simple model of Galactic rotation fails to reproduce the observations, even if we
assume $\Theta_0 = 220$ or $ 260 \, \text{km s}^{-1}$ (upper and lower dash-dotted black lines, respectively). We also
tried modifying the $(U_{\odot},V_{\odot},W_{\odot})$ components of the solar motion
(equivalent to adding a systematic motion to the LSR), but without satisfactory
results. We finally obtained a satisfactory fit by assigning to the stars associated with the Local Arm an additional
systematic velocity of $\Delta V_C=6 \, \text{km s}^{-1}$ in the direction of Galactic rotation and $\Delta V_R=1 \,
\text{km s}^{-1}$ in the radial direction. Such a systematic velocity could be inherited from the gas from which they
were born, which will deviate from pure rotation about the galactic center thanks to post-shock induced motions
associated with the Local Arm feature. Similar, but different, systematics may be at play for the other major
spiral arms, which we have not tried to model given the limited volume that is sampled by this \emph{Hipparcos} derived
dataset. In any case, the addition (or not) of these systematic motions parallel to the Galactic plane does not
significantly influence the the proper motions in galactic latitude, as discussed in Section \ref{results}.

As is well known, the International Celestial Reference Frame (ICRF) is the practical materialization of the
International Celestial Reference System (ICRS) and it is realized in the radio frequency bands, with axes intended to
be fixed with respect to an extragalactic intertial reference frame. The optical realization of the ICRS is based on
\emph{Hipparcos} catalogue and is called \emph{Hipparcos} Celestial Reference Frame (HCRF). \cite{VanLeeuwen2007} found that the
reference frame of the new reduction of \emph{Hipparcos} catalogue was identical to the 1997 one, aligned with the ICRF within
0.6 mas in the orientation vector (all 3 components) and within 0.25 mas/yr in the spin vector $\omega$ (all 3
components) at the epoch 1991.25. It is evident that a non-zero residual spin of the HCRF with respect to the ICRF
introduces a systematic error in the \emph{Hipparcos} proper motions. Depending on the orientation and on the magnitude of the
spin vector, the associated systematic proper motions can interfere or amplify a warp signature and must therefore be
investigated and taken into account \citep{Abedi:2015, Bobylev:2010, Bobylev:2015}. 
In the following section, when modelling the HIP2 sample, we consider the effects of such
a possible spin, adding the resulting systematic proper motions to the simulated catalogues following Equation 18 of
\cite{Lindegren:1995}, and using the residual spin vector $(\omega_x,\omega_y,\omega_z) \simeq (-0.126,+0.185,+0.076)
\, \text{mas} \, \text{yr}^{-1}$
 as measured by \emph{Gaia} \citep{Lindegren:2016}.

\subsection{Error model \label{SecErr}}

Our approach to confronting models with observations is to perform this comparison in the space of the observations.
Fundamental to this approach is having a proper description (i.e. model) of the uncertainties in the data. For this
purpose we construct an empirical model of the astrometric uncertainties in our two samples from the two catalogues
themselves. Below we first describe the astrometric error model for the HIP2 sample, based on the errors in the HIP2
catalogue, and then that of the TGAS(HIP2) sample based on the error properties of the \emph{Hipparcos} subsample in \emph{Gaia} DR1.
We note that, while we are here principally interested in the proper motions, we must also model the uncertainties of
the observed parallaxes $\varpi$ since we have applied the selection criteria $\varpi < 2$ mas to arrive at our OB
samples, and this same selection criteria must therefore be applied to any synthetic catalogue to be compared to our
sample.

\subsubsection{\emph{Hipparcos} error model}
\label{HipErr}

It is known that the \emph{Hipparcos} astrometric uncertainties mainly depend on the apparent magnitude (i.e. the S/N of the
individual observations) and on the ecliptic latitude as a result of the scanning law of the \emph{Hipparcos} satellite, which
determined the number of times a given star in a particular direction on the sky was observed. These dependencies are
not quantified in \cite{VanLeeuwen2007}, which only reports the {\em formal} astrometric uncertainties for each star.
To find the mean error of a particular astrometric quantity as a function of apparent magnitude and ecliptic latitude
we selected the stars with $(B-V) < 0.5$ from the HIP2 catalogue, consistent with the color range of our selected
sample of OB stars. We then bin this sample with respect to apparent magnitude and ecliptic latitude
and find, for each bin, the median errors for right ascension $\alpha$, declination $\delta$, parallax $\varpi$,
proper motions $\mu_{\alpha*}$ and $\mu_{\delta}$.
The resulting tables are
reported in the Appendix, which gives further details on their construction.

\begin{figure}[ht]
\resizebox{\hsize}{!}{\includegraphics{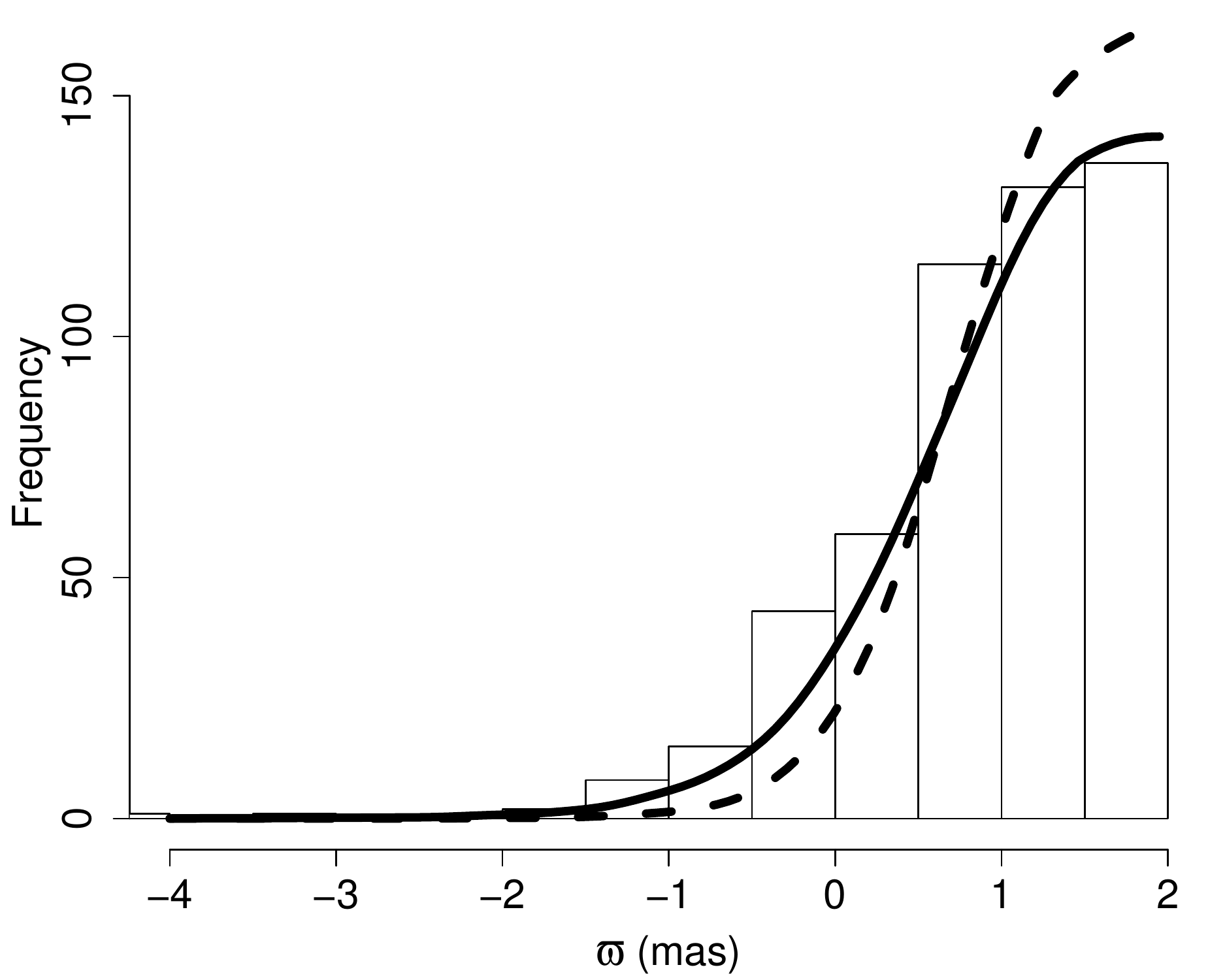}}
\caption{The histogram shows the observed parallax distribution. The dashed and the solid curves show, respectively, the synthetic distributions with F=1 and 1.5 (see text for explanation). }
\label{Plx_distr}
\end{figure}

However, before using these formal HIP2 uncertainties to generate random errors for our simulated stars, we first
evaluate whether the formal errors adequately describe the actual accuracy of the astrometric quantities. For this
purpose the distribution of observed parallaxes is most useful, and in particular the tail of the negative parallaxes,
which is a consequence of the uncertainties since the true parallax is greater than zero. In fact, using the mean
formal uncertainties in the parallax to generate random errors, we are unable to reproduce the observed parallax
distribution in our sample (see Figure \ref{Plx_distr}). Assuming that our model correctly describes the true
underlying distance distribution, we find that the formal HIP2 uncertainties must be inflated by a factor of $F=1.5$ to
satisfactorily reproduce the observed distribution. This factor $F$ is then also applied to the mean formal
uncertainties of the other astrometric quantities. We note that this correction factor is larger than that implied from
an analysis of the differences between the \emph{Hipparcos} and \emph{Gaia} DR1 parallaxes. (See Appendix B of \cite{Lindegren:2016}.)

To better fit the HIP2 proper motion distributions we also take into consideration stellar binarity. Indeed,
approximately $f_b \approx 20 \%$ of stars of our sample has been labelled as binary, either resolved or unresolved, in
the HIP2 catalogue. For these stars, the uncertainties are greater than for single stars. Therefore, we inflate the
proper motion errors for a random selection of $20\%$ of our simulated stars by a factor of $f_{bin}=1.7$ to arrive at
a distribution in the errors comparable to the observed one.

Finally, we also performed similar statistics on the correlations in the HIP2 astrometric quantities published by
\cite{VanLeeuwen2007}, using the four elements of the covariance matrix relative to the proper motions (see Appendix B
of \cite{Michalik:2014}). We find that the absolute median correlations are less than 0.1, and therefore we do not take
them into account.

\subsubsection{TGAS(HIP2) error model}

\begin{figure}[ht]
\resizebox{\hsize}{!}{\includegraphics{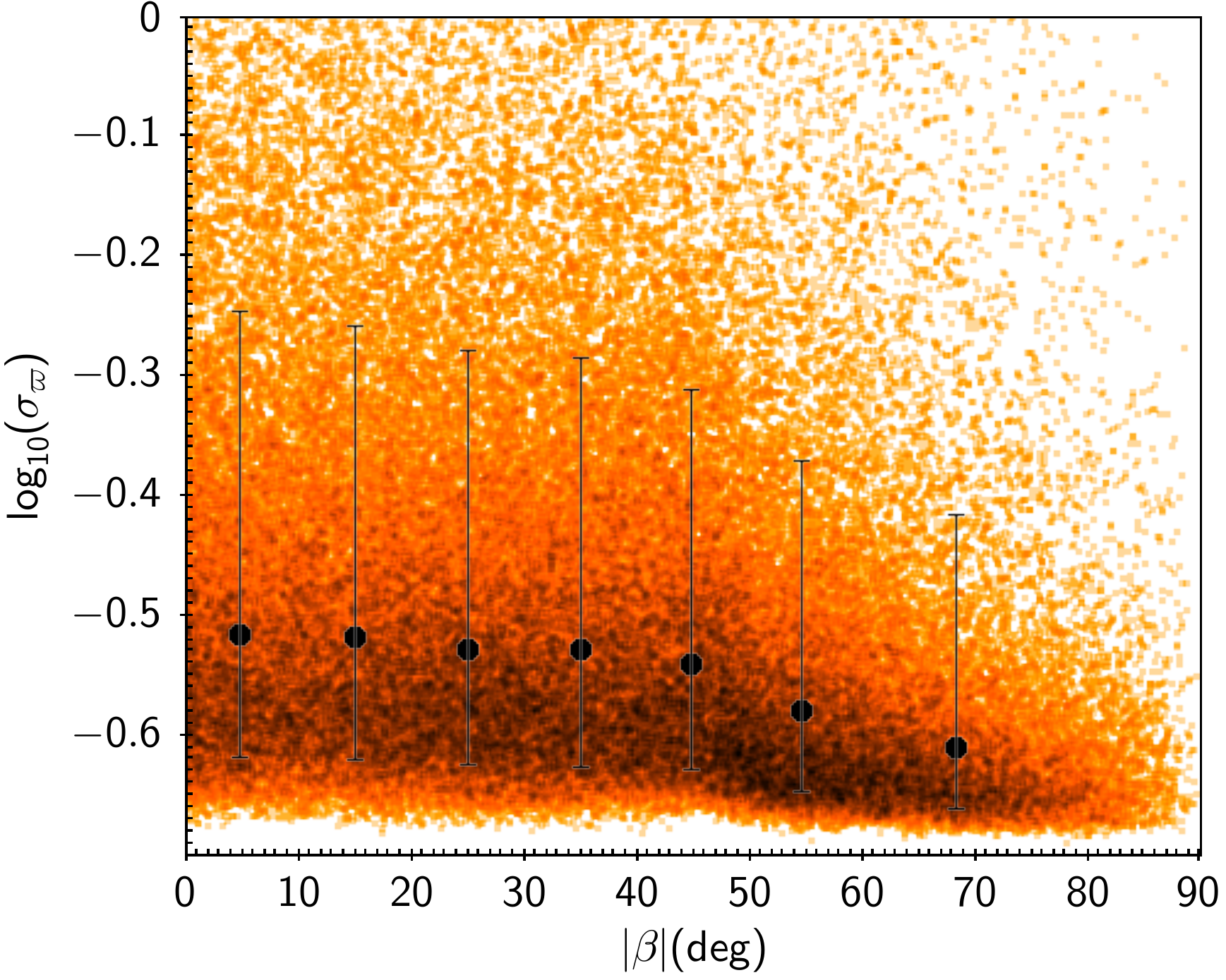}}
\caption{ Logarithm of the parallax errors {\bf (mas)} in
function of ecliptic latitude for the \emph{Hipparcos} subsample in TGAS. The point show the medians, while the error bars
show the $10^{th}$ and the $90^{th}$ percentiles of the distribution. }
\label{plaxerr_vs_beta}
\end{figure}

\begin{figure}[ht]
\resizebox{\hsize}{!}{\includegraphics{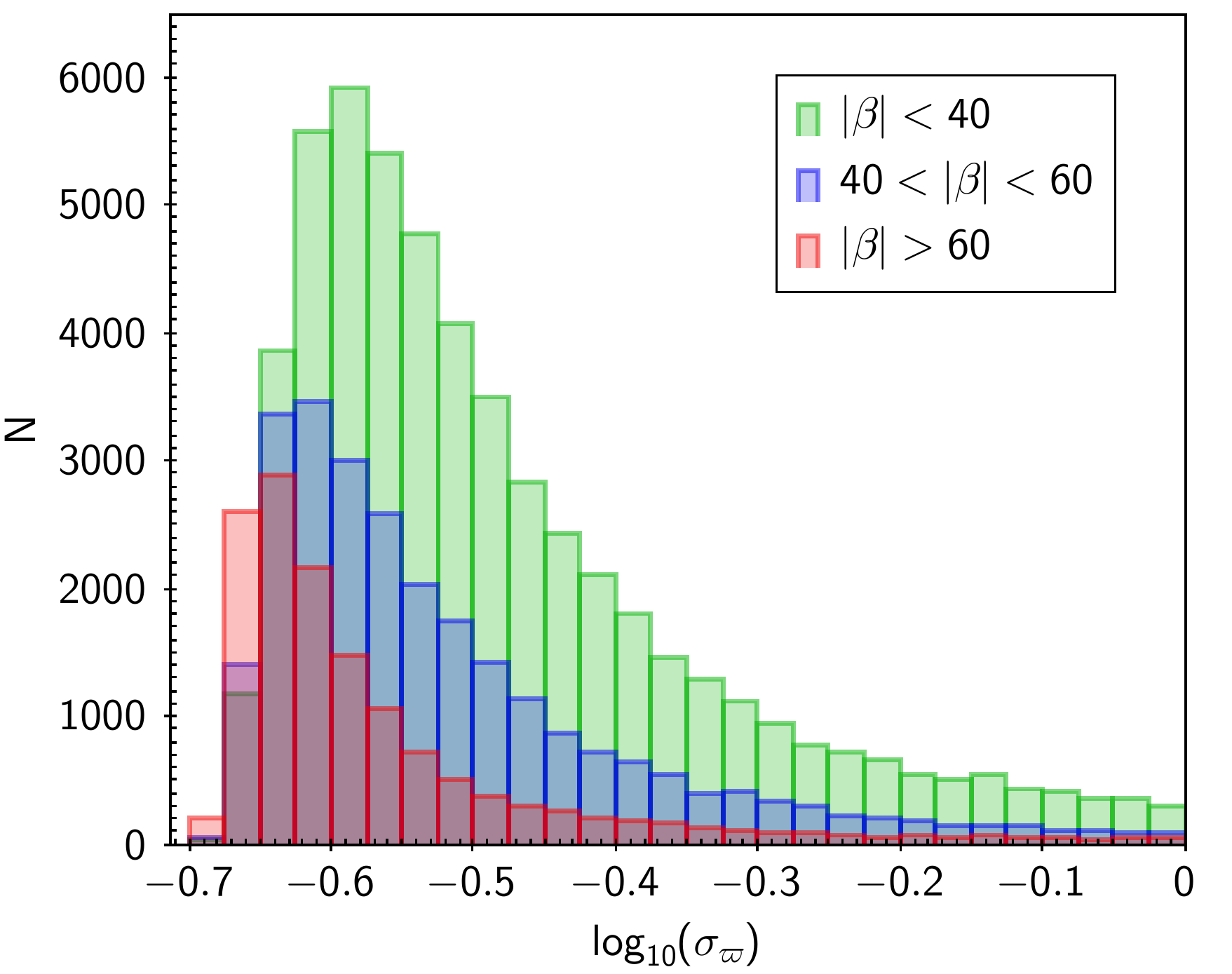}} \caption{ Distribution of the logarithm of the parallax uncertainties (mas) for the HIP-TGAS stars. Three subsets with different ecliptic latitude are shown.} \label{plxerr_dist}
\end{figure}

A detailed description of the astrometric error properties of the TGAS subset in \emph{Gaia} DR1 is described in
\cite{Lindegren:2016}. However, on further investigation we found that the error properties of the subset of 93635
\emph{Hipparcos} stars in \emph{Gaia} DR1 are significantly different with respect to the larger TGAS sample. In
particular, we find that the zonal variations of the median uncertainties seen with respect to position on the sky are
much less prominent for the \emph{Hipparcos} stars in DR1, and are only weakly dependent on ecliptic latitude. The
parallax errors with respect to ecliptic latitude are shown in Figure \ref{plaxerr_vs_beta}. Meanwhile the TGAS(HIP2)
parallax errors show no apparent correlation with respect to magnitude or color. Figure \ref{plxerr_dist} shows the
distribution of parallax uncertainties for three ecliptic latitude bins, which we model with a gamma distribution
having the parameters reported in Table \ref{tab:gamma}. However, \cite{Lindegren:2016} has warned that there is
an additional systematic error in the parallaxes at the level of 0.3mas. In this work we only use the parallaxes to
split our sample in two subsets, and we have verified that adding an additional random error of 0.3 mas to account for
these possible systematic errors does not affect our results.
\begin{table}[ht]
   \caption{ The shape parameter $k$ and the scale parameter $\theta$ of the gamma distributions used to model the
    $\log_{10} (\sigma_{\varpi})$ distributions shown in Figure \ref{plaxerr_vs_beta}). An additional offset is required in order to fit the distributions. }
  \label{tab:gamma}
  \centering
  \begin{tabular}{c c c c}
    \hline
    \hline
     Ecliptic latitude $| \beta |$ (deg) &  $k$ & $\theta$ & offset\\
    \hline
     0-40 & 1.5 & 0.113 & -0.658\\
     40-60 & 1.2 & 0.115 & -0.658\\
     60-90 & 1.1 & 0.08 & -0.67\\
     \hline
  \end{tabular}
\end{table}

\begin{figure}[ht]
\resizebox{\hsize}{!}{\includegraphics{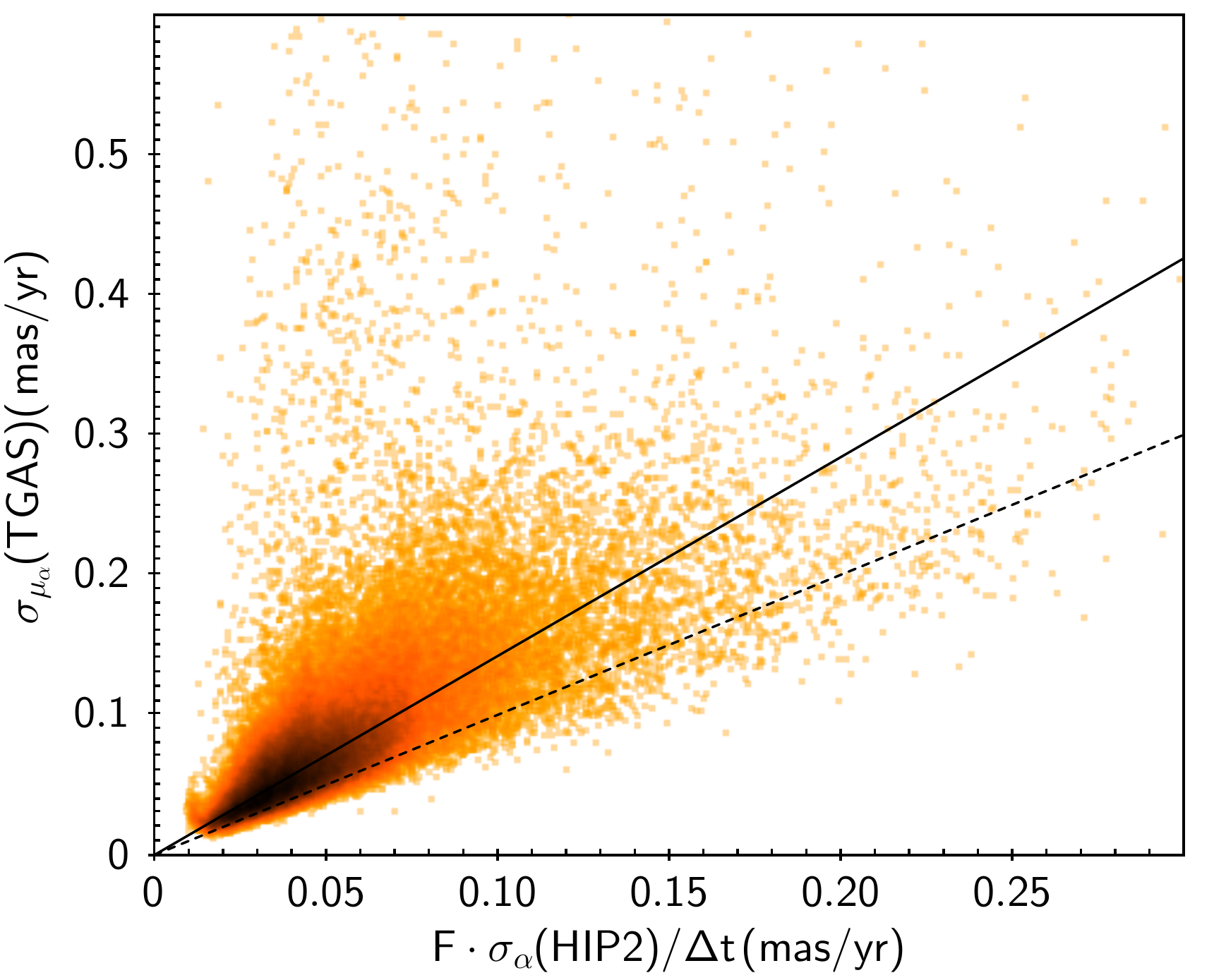}}
\caption{ For each star of the \emph{Hipparcos} subset in TGAS, the published error $\sigma_{\mu_{\alpha*}}$(TGAS) is compared to the prediction based on \emph{Hipparcos} uncertainties $F \, \sigma_\alpha^H(m,\beta) / \Delta t$ (see text). The dashed line represents the bisector. The solid line has null interceipt and coefficient $C_\alpha=1.42$, which is used to calibrate our error model (see text).}
\label{TGASpm_errors}
\end{figure}

The errors for the proper motions also show a weak dependence on ecliptic latitude, as well as additional dependence
with respect to magnitude. Indeed, we find that the proper motion errors for the \emph{Hipparcos} subset in \emph{Gaia} DR1 are
strongly correlated the \emph{Hipparcos} positional errors, as one would expect, given that the \emph{Hipparcos} positions are used
to constrain the \emph{Gaia} DR1 astrometric solutions \citep{Michalik:2015}. We use this correlation to model the proper motion errors of
the \emph{Hipparcos} subsample in DR1. Figure \ref{TGASpm_errors} show the agreement which results when we take as our model
$\sigma_{\mu_\alpha} = C_\alpha \, \bigl[ F \, \sigma_\alpha^H(m,\beta) / \Delta t \bigr]$ , where F is the correction factor
 applied to the \emph{Hipparcos} astrometric uncertainties, as described in Section \ref{HipErr}, $\sigma_\alpha^H(m,\beta)$ is the \emph{Hipparcos} error in right
ascension, interpolated from table \ref{table:erralpha2007} in the Appendix, and $\Delta t$ is the difference between the \emph{Gaia} (J2015)
and \emph{Hipparcos} (J1991.25) epoch. The adopted coefficient $C_\alpha=1.42$ is the median of  $\sigma_{\mu_\alpha} / \bigl[ F \, \sigma_\alpha^H(m,\beta) / \Delta t \bigr]$ for the stars of the TGAS(HIP2) sample. An analogous model is used for $\sigma_{\mu_\delta}$, with $C_\delta = 1.44$.

\begin{figure}[ht]
\resizebox{\hsize}{!}{\includegraphics{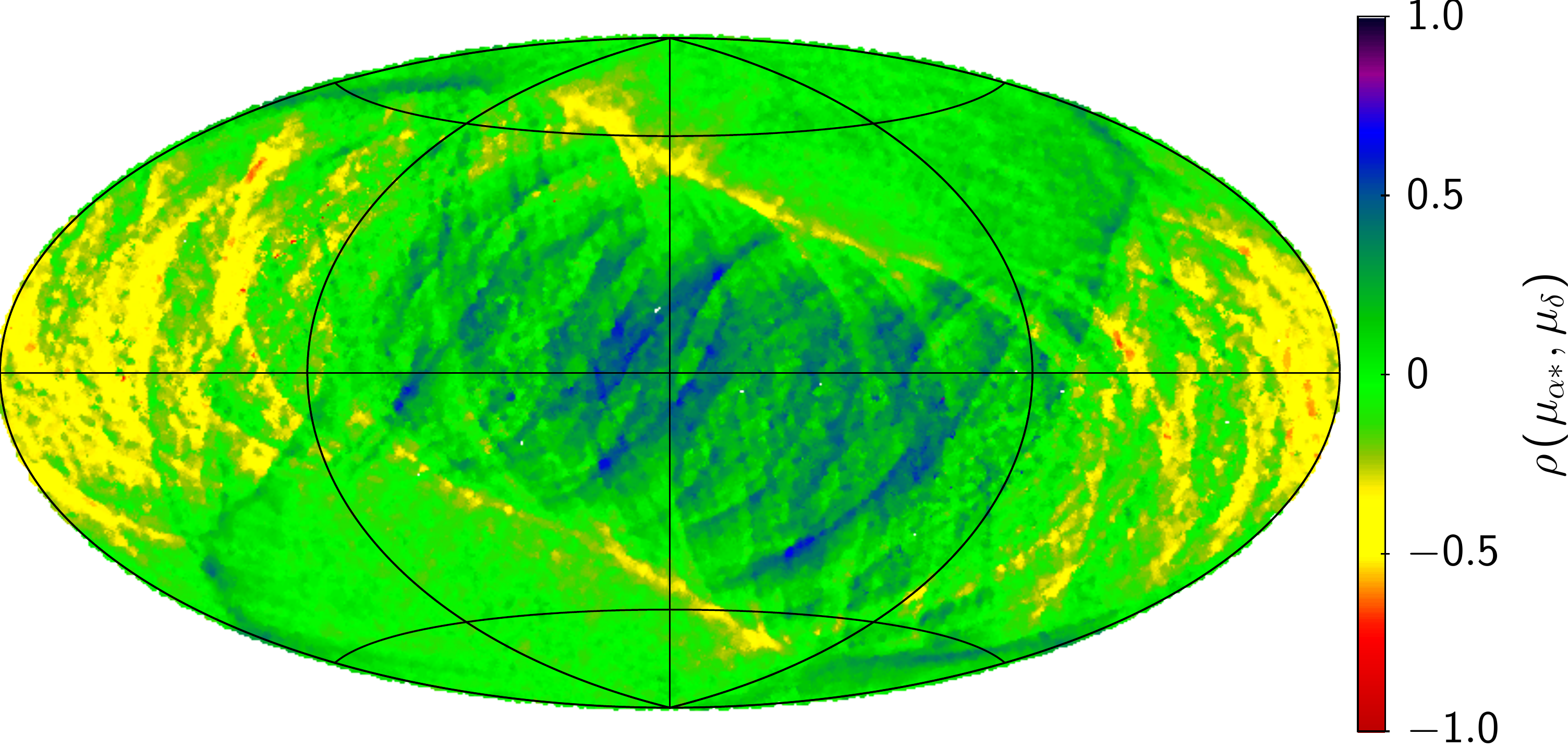}}
\resizebox{\hsize}{!}{\includegraphics{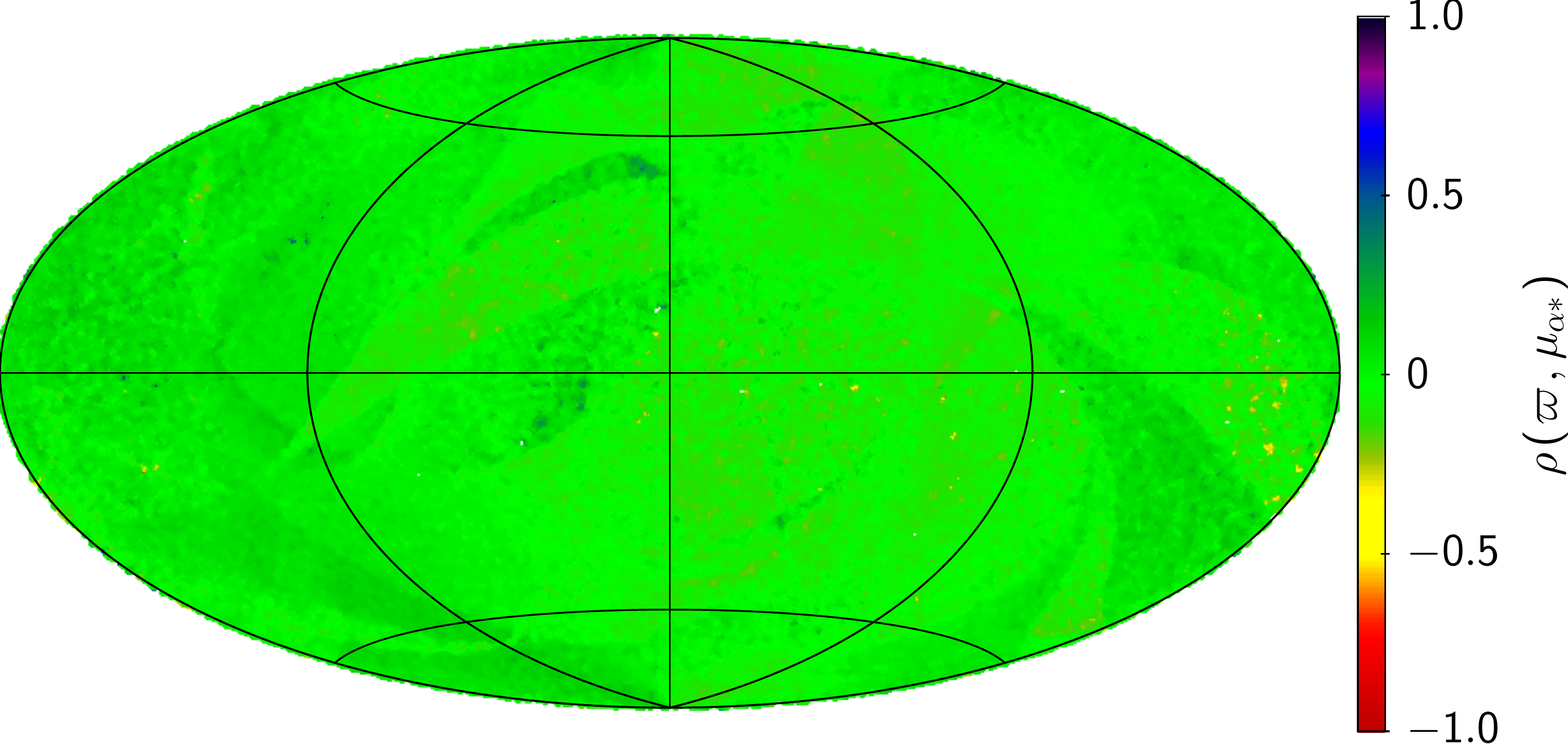}}
\resizebox{\hsize}{!}{\includegraphics{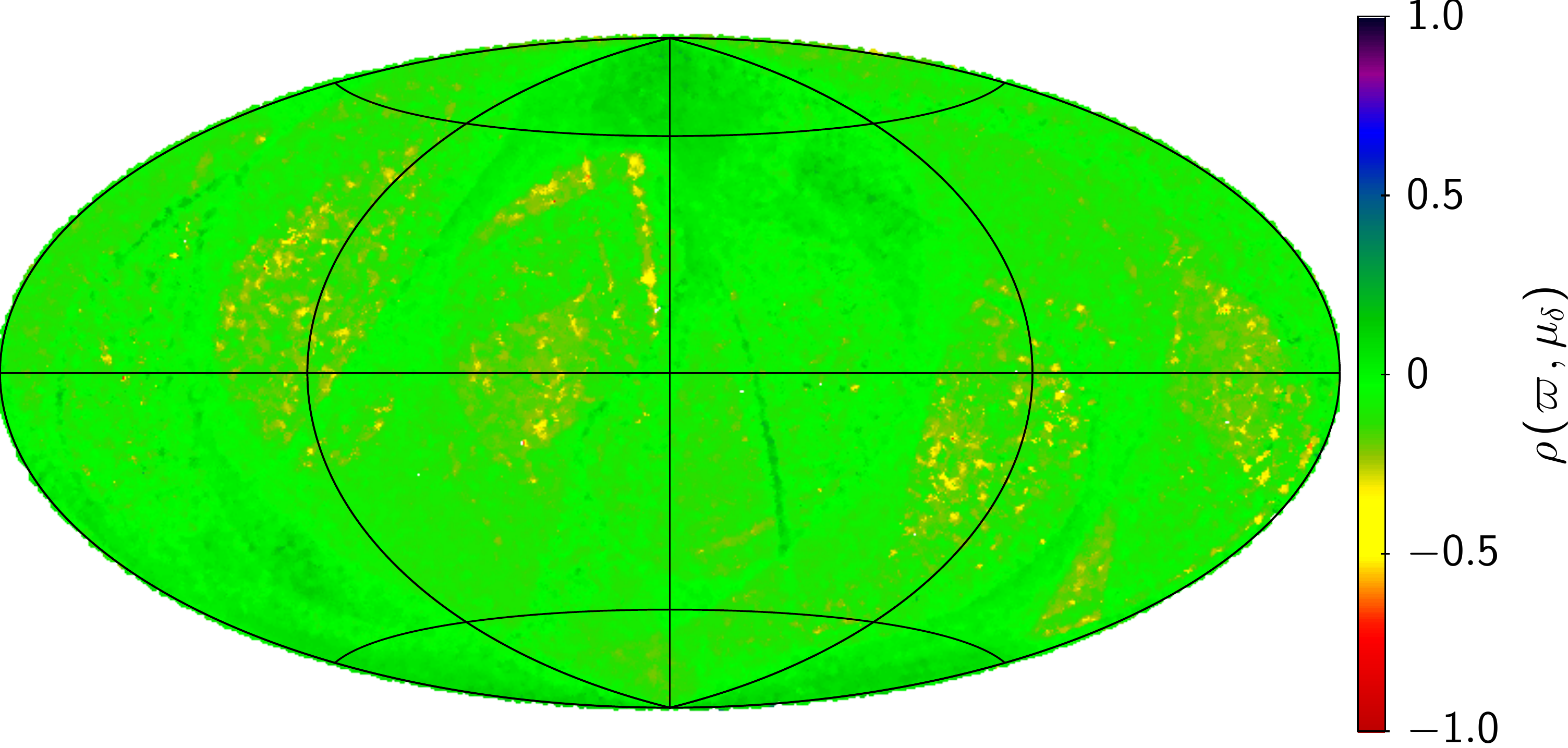}}
\caption{ Median correlations between parallaxes $\varpi$ and proper motions $\mu_\alpha$, $\mu_\delta$ in HIP-TGAS. } \label{corr_HIP_TGAS}
\end{figure}

Finally, in contrast to the correlations in the HIP2 sample, we find that the correlations in DR1 between the
astrometric quantities of the \emph{Hipparcos} subsample vary strongly accross the sky, but are significantly different from
the complete TGAS sample, shown in Figure 7 of \cite{Lindegren:2016}. Figure \ref{corr_HIP_TGAS} shows the variation
across the sky of the correlations between the parallaxes and the proper motions. As we can see, the correlations between
proper motion components are relevant. To take this into account, we generate the synthetic proper motion errors from a bivariate gaussian distribution with a covariance matrix which includes the $\sigma_{\mu_\alpha}$ and $\sigma_{\mu_\delta}$ predicted by the above described model and the correlations from the first map in Figure \ref{corr_HIP_TGAS}.

\section{Comparison between models and data}
\label{results}

In this section we first attempt to derive the systematic vertical motions from the observed proper motions and compare
these with those predicted from the model, to demonstrate the weaknesses of this approach. We then compare the observed
proper motions of our two samples with the proper motion distributions derived from models with and without a warp.

\subsection{Mean vertical velocity in function of Galactocentric radius}
\label{SecRvz}

\begin{figure}[ht]
\resizebox{\hsize}{!}{\includegraphics{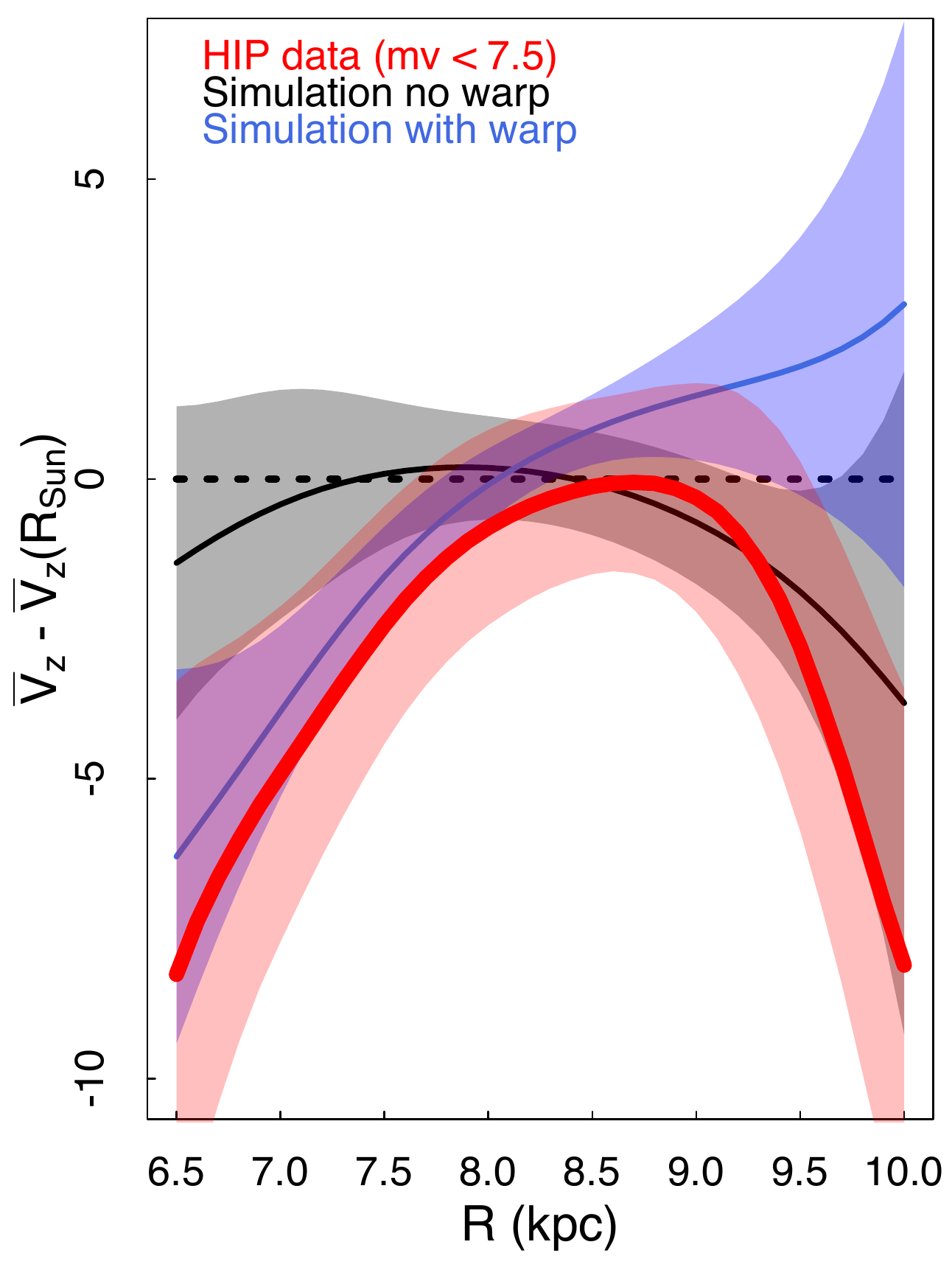}} \caption{Bivariate Nadaraya-Watson regression estimator of
stellar vertical velocities as a function of Galactocentric radius, using a bandwidth of $h=0.5 \,\text{kpc}$. The same regression
bandwidth has been used for the data (red), the no-warp model (black) and warp model of \cite{Yusifov:2004} (blue).
The 95 $\%$
bootstrap confidence band is shown for the data. For each of the two models, the non parametric regressions are
performed for 20 simulated catalogues, obtaining the curves as the mean values and the shaded areas as the 95 $\%$
uncertainty.  } \label{Figrvz}
\end{figure}

We first consider the mean vertical velocity $ \overline{v}_z$ as a function of Galactocentic radius $R$ that one would
derive from the measured proper motions and spectro-photometric distances, comparing the data with what we expect from
the no-warp and warp models. In the no-warp case, the true mean vertical velocities are zero, while a warp model
predicts that they increase with $R$ outside the radius $R_w$ at which the Galactic warp starts (see equation
\ref{zwarp}.) Figure \ref{Figrvz} shows the mean vertical velocities, after removing the solar motion, for the data and
the no-warp and warp simulated catalogues.
The simulated catalogues include the modelled errors, as described in Section \ref{SecErr}. Taking into account both
distance and proper motion errors, the observed trend is biased toward negative velocities with increasing distance.
This bias is particularly evident with the no-warp model, where the true $\overline{v}_z(R)=0$ (dashed line in Figure
\ref{Figrvz}), but similarly affects the warp model. One might be tempted to proceed to compare models to the data in
this space of derived quantities, assuming the error models are correct, but this approach gives the most weight to the
data at large distances, i.e. those with the highest errors and the most bias. Indeed, from Figure \ref{Figrvz} one
might quickly conclude that the data was consistent with the no-warp model, based however on trends that are dominated
by a bias in the derived quantities.

A better approach is to compare the data to the models in the space of the observations, i.e. the mean proper motions
as a function of position on the sky, thereby avoiding the biases introduced by the highly uncertain distances. That
is, it is better to pose the question: which model best reproduces the observations? Our approach will make
minimum use of the spectro-photometric distances and parallaxes to avoid the strong biases introduced when using
distances with relatively large uncertainties to arrive at other derived quantities, as in the example above. Indeed,
whether based on spectrophotometric data or parallaxes, distance is itself a derived quantity that can suffer from
strong biases \citep{BailerJones:2015}. Nevertheless, distance information is useful. We have already used the distance
information contained in the parallaxes for selecting our two samples with the criteria of $\varpi < 2$ mas (see
Section \ref{SecData}), while in Section \ref{Seclmub}, we will split our HIP and TGAS(HIP) samples into nearby and
distant sources.

\subsection{Proper motion $\mu_b$ in function of Galactic longitude \label{Seclmub}}

\begin{figure*}[ht]
\centering
\includegraphics[scale=0.5]{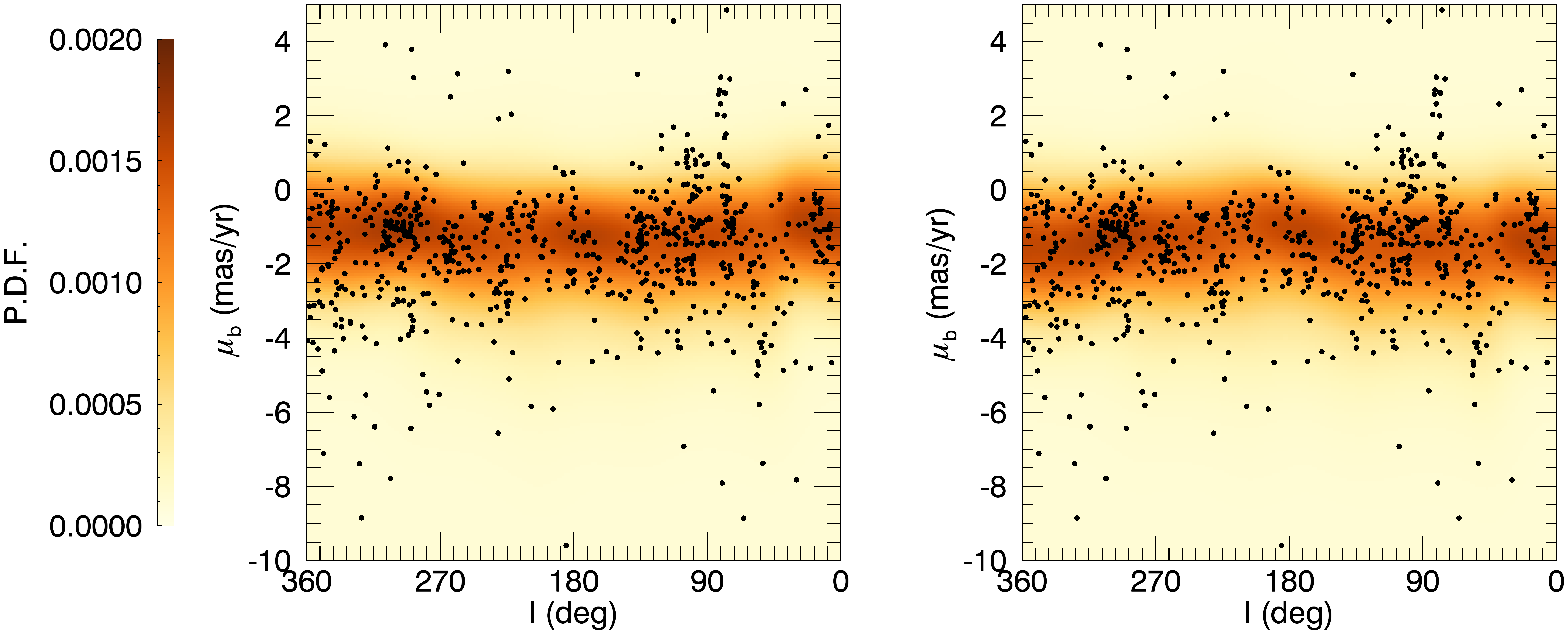} \caption{Distribution of the
TGAS-HIP data in the $l$-$\mu_b$ plane (black dots), together with the Probability Density Function $P(\mu_{b}|l)$
predicted by the no-warp model (left panel) and the warp model of \cite{Yusifov:2004} (right panel). } \label{lkhd_tgas}
\end{figure*}

The aim of the present work is to determine whether a warp is favored over a no-warp model in either of our
\emph{Hipparcos} or \emph {Gaia} DR1 samples. In Section \ref{SecWarp} we showed that the Galactic warp, if stable,
results in a distinct trend in the proper motions perpendicular to the galactic plane $\mu_b$ with respect to galactic
longitude, with higher (i.e. more positive) proper motions toward the Galactic anti-center (see Figure \ref{mubth}).
Our approach is to compare the distribution of the observed proper motions with the expectations derived from a warp
model and a no-warp model, taking into full account the known properties of the astrometric errors. To achieve this we
adopt the approach of calculating the likelihood associated with a given model as the probability of the observed data
set arising from the hypothetical model \citep[as described in][]{Peacock:1983}.

Given an assumed model (i.e. parameter set), we generate 500 thousand stars and perform a two-dimensional kernel
density estimation in $l - \mu_b$ space to derive the conditional Probability Density Function (PDF), $P(\mu_b|l)$, the
probability of observing a star with proper motion $\mu_b$ at a given galactic longitude, where $P(\mu_b|l)$ is
constrained to satisfy $\int P(\mu_b|l) \, \text{d} \mu_b = 1$. The motivation for using $P(\mu_b|l)$ is that we want
to assign the probability of observing a given value of $\mu_b$ independent of the longitude distribution of the stars,
which is highly heterogeneous. That is, we wish to quantify which model best reproduces the observed trend of the
proper motions with respect to longitude. In any case, we also performed the below analysis using $P(l,\mu_b)$,
imposing the normalization $\int{ P(l,\mu_b) \, \text{d}}l \text{d} \mu_b = 1$, and obtain similar results. Figure
\ref{lkhd_tgas} shows the conditional PDFs for the warp/no-warp models for the TGAS(HIP2) sample. The PDFs for the HIP2
sample (not shown) are very similar, as the proper motion distribution is dominated by the intrinsic velocity
dispersion of our sample rather than by the proper motion errors.

Once the PDFs for the two different models are constructed, we found the probability $P(\mu_{b,i}|l_i)$ associated with
each $i$-th observed star according to each PDF. The likelihood associated with the model is $ L = \prod_{i=1}^N
P(\mu_{b,i}|l_i)$, where N is the total number of stars in our dataset; for computational reasons, we used instead the
log-likelihood $\ell = ln(L) = \sum_{i=1}^N ln(P(\mu_{b,i}|l_i)$.  Also for practical reasons we applied a cut in
$\mu_b$, considering only the range $( -10 < \mu_b < 5)$ mas/yr when calculating $\ell$, reducing our HIP2 dataset to
989 stars, and our TGAS(HIP2) sample to 791 stars. 
Below we will confirm that this clipping of the data does not impact our results by
considering alternative cuts on $\mu_b$.

For a given sample, the difference between the log-likelihoods of a warp model and the no-warp model (i.e. the
ratio of the likelihoods), $\Delta \equiv \ell_{WARP} - \ell_{NOWARP}$, is found as a measure of which model is more
likely. We performed a bootstrap analysis of the log-likelihood to quantify the significance level of the obtained $\Delta$.
Bootstrap catalogues were generated by randomly extracting stars N times from the observed set of N stars of the dataset
(resampling with replacement). As suggested by \cite{Feigelson:2012}, $N_B \approx N (\ln N)^2$ bootstrap resamples
were generated. For each bootstrap resample, the log-likelihood was computed for the two models and the
log-likelihood difference $\Delta$ was calculated. Finally, after $N_B$ resamples, the standard deviation
$\sigma_{\Delta}$ of the distribution of $\Delta$ is determined, while the integral of the normalized $\Delta$
distribution for $\Delta
> 0$ gives $P(\Delta > 0)$, the probability of the warp model being favoured over the no-warp model.

\begin{table*}[ht]
   \caption{
   Difference of the log-likelihoods of the warp and nowarp models $\Delta \equiv \ell_{WARP} - \ell_{NOWARP}$ according to
   the warp parameters reported in \cite{Drimmel:2001} (dust and stellar model) and \cite{Yusifov:2004}. Log-likelihoods are calculated
   with the TGAS(HIP2) sample ($\varpi < 2$ mas), containing 758 stars. We also show the results for the nearby ($(1 < \varpi < 2)$ mas, 296 stars)
   and for the distant objects ($\varpi < 1$ mas, 462 stars). The standard deviation $\sigma_{\Delta} $ and the probability $P (\Delta>0)$ are calculated using bootstrap resamples (see text).}
  \label{tab:likelihood_TGAS_warpmod}
  \centering
  \begin{tabular}{l c c c c c c c c c}
    \hline
    \hline
       & \multicolumn{3}{c}{$\varpi < 2$ mas} & \multicolumn{3}{c}{ $(1 < \varpi < 2)$ mas} & \multicolumn{3}{c}{ $\varpi < 1$ mas}\\
       \cmidrule(lr){2-4} \cmidrule(lr){5-7} \cmidrule(lr){8-10}
    Warp model   &  $\Delta $ &  $\sigma_{\Delta} $ & $P (\Delta>0)$ &  $\Delta $ & $\sigma_{\Delta} $ & $P (\Delta>0)$  &  $\Delta $ &  $\sigma_{\Delta} $ & $P (\Delta>0)$\\
    \hline
     \cite{Drimmel:2001}, dust             & -41.24 & 9.72 & 0.00                   & -2.94 & 5.04 & 0.28  & -60.35 & 10.90 & 0.00 \\
     \cite{Drimmel:2001}, stars            & -5.47 & 3.93 & 0.09                    & 3.13 & 1.92 & 0.95     & -22.54 & 16.79 & 0.04 \\
     \cite{Yusifov:2004}                      & -2.69 & 6.99 & 0.35                    & 13.93 & 3.47 & 1.00   &  -10.48 & 25.81 & 0.32 \\
     \hline
  \end{tabular}
\end{table*}

\begin{table*}[ht]
   \caption{ Difference of the log-likelihoods of the warp \citep{Yusifov:2004} and nowarp models $\Delta \equiv \ell_{WARP} - \ell_{NOWARP}$
   for the TGAS(HIP2) sample. Results are shown for the whole sample (\emph{All}, $m_v < 8.5$) and for the bright sample ($m_v < 7.5$).
   We also report the results obtained removing the objects with high $\Delta Q$, where $\Delta Q$
   is the difference between the TGAS and \emph{Hipparcos} proper motion \citep{Lindegren:2016}. $\Delta Q_{95\%}=23$ and
   $\Delta Q_{90\%}=11$ are the percentiles the $\Delta Q$ distribution for all the \emph{Hipparcos} subset in TGAS.
   We also present the results excluding the stars labelled as binaries in \cite{VanLeeuwen2007}.
   The $95 \%$ ($90 \%$) Confidence Interval is obtained considering the stars with proper motions
   $\mu_b$ between the $2.5^{th}$ and the $97.5^{th}$ (the $5^{th}$ and the $95^{th}$) percentiles of the
   whole $\mu_b$ distribution (with 767 stars), without restricting to the range $( -10 < \mu_b < 5)$ mas/yr (see text).
   The standard deviation $\sigma_{\Delta} $ and the probability $P (\Delta>0)$ are calculated using bootstrap resamples (see text).
 }
  \label{tab:likelihood_TGAS}
  \centering
  \begin{tabular}{l c c c c c c c c c c c c}
    \hline
    \hline
       & \multicolumn{4}{c}{$\varpi < 2$ mas} & \multicolumn{4}{c}{ $(1 < \varpi < 2)$ mas} & \multicolumn{4}{c}{ $\varpi < 1$ mas}\\
       \cmidrule(lr){2-5} \cmidrule(lr){6-9} \cmidrule(lr){10-13}
    sample  & $N_{stars}$ &  $\Delta $ &  $\sigma_{\Delta} $ & $P (\Delta>0)$ & $N_{stars}$ &  $\Delta $ & $\sigma_{\Delta} $ & $P (\Delta>0)$ & $N_{stars}$ &  $\Delta $ &  $\sigma_{\Delta} $ & $P (\Delta>0)$\\
    \hline
      All                                              & 758 & -2.69 & 6.99 & 0.35                        & 296 & 13.93 & 3.47 & 1.00 & 462 & -10.48 & 25.81 & 0.32 \\
    $m_v < 7.5$                                & 310 & 0.68 & 4.44 & 0.57                           &  129 & 4.59 & 2.09 & 0.99 & 181  & -0.47 & 4.03 & 0.45 \\
     $\Delta Q < \Delta Q_{95\%}$ & 749 & -3.73 & 6.90 & 0.30                         & 293 & 13.86 & 3.50 & 1.00 & 456 & -11.54 & 25.89 & 0.31 \\
     $\Delta Q < \Delta Q_{90\%}$ & 690 & -7.35 & 6.56 & 0.13                          & 257 & 10.97 & 3.07 & 1.00 & 433 & -12.18 & 25.93 & 0.30 \\
     No HIP2 binaries                       & 672 & -2.75 & 6.81 & 0.34                          & 267 & 15.03 & 3.40 & 1.00 & 405  & -11.85 & 26.27 & 0.31 \\
     $95 \%$ Conf. Int.                    & 730 & -0.86 & 5.94 & 0.44                         & 289 &  10.30 & 3.23 & 1.00 & 444 & -13.99 & 5.23 & 0.00   \\
     $90 \%$ Conf. Int.                    & 692 & 2.79 & 4.91 & 0.72                          & 273 & 9.11 & 2.71 & 1.00 & 420 & -10.85 & 4.13 & 0.00  \\
     \hline
  \end{tabular}
\end{table*}

\begin{table*}[ht]
   \caption{ Difference of the log-likelihoods of the warp \citep{Yusifov:2004} and nowarp models $\Delta \equiv \ell_{WARP} - \ell_{NOWARP}$ for the HIP2 sample. Results are shown for the whole sample (\emph{All}, $m_v < 8.5$) and for the bright sample ($m_v < 7.5$). We also present the results obtained removing the stars labelled as binaries in \cite{VanLeeuwen2007}. The standard deviation $\sigma_{\Delta} $ and the probability $P (\Delta>0)$ are calculated using bootstrap resamples (see text). }
   \label{tab:likelihood_HIP2}
  \centering
  \begin{tabular}{l c c c c}
    \hline
    \hline
    HIP2 subset & $N_{stars}$ &  $\Delta $ &  $\sigma_{\Delta} $ & $P (\Delta>0)$  \\
    \hline
      All  & 989 & -14.99 & 6.50 & 0.01 \\
      All No HIP2 binaries  & 838 & -11.86 & 5.74 & 0.02 \\
      $m_v < 7.5$  & 498 & -1.68 & 4.47 & 0.35 \\
      $m_v < 7.5$ No HIP2 binaries  & 404 & -2.29 & 4.21 & 0.29 \\
     \hline
  \end{tabular}
\end{table*}

Table \ref{tab:likelihood_TGAS_warpmod} collects the results for the TGAS(HIP2) dataset for the three different
warp models whose parameters are given in Table \ref{tab:warp_models}, for the full dataset as well as for the two
subsets of distant ($\varpi < 1$ mas) and nearby ($ 2 > \varpi > 1$ mas) stars. For the full dataset none of the warp
models are favored over a no-warp model, though the model of \cite{Yusifov:2004} cannot be excluded. However, on splitting our sample into
distant and nearby subsamples, we find some of the warp models are clearly favoured over the no-warp model for the
nearby stars.

In Tables \ref{tab:likelihood_TGAS} and \ref{tab:likelihood_HIP2} we show the results of our analysis, for both the
HIP2 and TGAS(HIP2) samples, considering various subsamples with alternative data selection criteria to test for
possible effects due to incompleteness and outliers. Separate PDFs were appropriately generated for the selections in
magnitude and parallax. Here we show the results using the \cite{Yusifov:2004} model, as it is the most consistent with
the data, as indicated by the the maximum likelihood measurements shown in Table \ref{tab:likelihood_TGAS_warpmod}.
Again, the chosen warp model is not clearly favored nor disfavored until we split our sample into distant and nearby
subsamples. 

Various selection criteria were applied to investigate the role of possible outliers in biasing the outcome.
We tried to remove the stars identified as binaries
in the \emph{Hipparcos} catalogue \citep{VanLeeuwen2007} and, for the TGAS subset,
the objects with a high difference between the \emph{Gaia} and \emph{Hipparcos} proper motion \citep{Lindegren:2016}.
We also removed the high-proper motion stars (i.e. the tails in the $\mu_b$ distributions),
to exclude possible runaway stars or nearby objects with significant peculiar motions.
As shown in Tables \ref{tab:likelihood_TGAS} and \ref{tab:likelihood_HIP2}, the exclusion of
these possible outliers doesn't change our findings, confirming that the warp model \citep{Yusifov:2004}
is preferred for the nearby objects, but rejected for the distant stars.

A further test was performed, removing the most obvious clumps in the $l,b$ and in the
$l,\mu_b$ space (for example the one centered on $l \approx 80 \degr$ and $\mu_b \approx 2.5$ mas/yr,
see Figure \ref{lkhd_tgas}), to study the effect of the intrinsic clumpiness of the OB stars. We
also removed the stars part of the known OB association Cen OB2 according to \cite{DeZeeuw:1999}.
The obtained results (here not shown) are very similar to the ones in Tables
\ref{tab:likelihood_TGAS} and \ref{tab:likelihood_HIP2}.


\section{Discussion}

We have used models of the distribution and kinematics of OB stars to find the expected distribution of proper
motions, including astrometric uncertainties, for two samples of spectroscopically identified OB stars from the New
\emph{Hipparcos} Reduction and \emph{Gaia} DR1. The resulting PDFs of the proper motions perpendicular to the galactic
plane produced by models with and without a warp are compared to the data via a likelihood analysis. We find that the
observed proper motions of the nearby stars are more consistent with models containing a kinematic warp signature than
a model without, while the more distant stars are not. Given that the warp signal in the proper motions is expected to
remain evident at large distances (see Section \ref{SecWarp}), this result is difficult to reconcile with the
hypothesis of a stable warp, and we are forced to consider alternative interpretations.

Keeping in mind that our sample of OB stars is tracing the gas, one possibility is that the warp in the gas starts well
beyond the Solar Circle, or that the warp amplitude is so small that no signal is detectable. However, most studies to
date suggest that the warp in the stars and in the dust starts inside or close to the Solar circle,
\citep{Drimmel:2001,Derriere:2001,Yusifov:2004,Momany:2006,Robin:2008}, while the warp amplitude in the gas at
R$\approx$10 kpc \citep{Levine:2006} is consistent with warp models of sufficiently small amplitudes. Indeed,
\cite{Momany:2006} found an excellent agreement between the warp in the stars, gas and dust using the warp model of
\cite{Yusifov:2004}, the same model that we used in Section \ref{Seclmub}. Another possible scenario is that the warp
of the Milky Way is a short-lived/transient feature, and that our model of a stable warp is not applicable. This
hypothesis would be consistent with the finding that the warp structure may not be the same for all Milky Way
components, as argued in \cite{Robin:2008}, but in contradiction to the findings of \cite{Momany:2006} cited above.
Finally, our expected kinematic signature from a stable warp could be overwhelmed, or masked, by other systematic
motions. Evidence has been found for vertical oscillations \citep{Widrow:2012, Xu:2015}, suggesting the presence of
vertical waves, as well kinematic evidence of internal breathing modes \citep{Williams:2013} in the disk. Both have
been attributed as being possibly caused by the passage of a satellite galaxy
\citep{Gomez:2013,Widrow:2014,Laporte:2016}, while breathing modes could also be caused by the bar and spiral arms
\citep{Monari:2015, Monari:2016}. If such effects as these are present, then sampling over a larger volume of the
Galactic disk will be necessary to disentangle the kinematic signature of the large-scale warp from these other
effects. Also, a comparison of the vertical motions of young stars (tracing the gas) and a dynamically old sample could
also confirm whether the gas might possess additional motions due to other effects. 

We have compared the proper motions of our two samples with the expectations from three warp models taken from the
literature. Among these the model based on the FIR dust emission \citep{Drimmel:2001} can be excluded based on the
kinematic data from \emph{Gaia} DR1 that we present here. In addition, the \cite{Drimmel:2001} dust warp model also
predicts a vertical motion of the LSR of 4.6 km s$^{-1}$, which would result in a vertical solar motion that clearly
inconsistent the measured proper motion of Sag A$^*$ \citep{Reid:2008}. This calls into question the finding of
\cite{Drimmel:2001} that the warp in the dust and stars are significantly different. However, the proper motion data
does not strongly favor the other two warp models, that of \cite{Yusifov:2004} and \cite{Drimmel:2001} based on the
stellar NIR emission: As pointed out in Section \ref{SecWarp}, the local kinematic signature produced by a warp model
with $\phi_w \neq 0$ is quite similar to that of a warp model with $\phi_w = 0$ of smaller amplitude. In short, the
parameters of even a simple symmetric warp cannot be constrained from local kinematics alone. In any case, we stress
that the observed kinematics of the most distant OB stars are not consistent with any of the warp models.

\section{Conclusion and future steps}
\label{Sec:concl}

Our search for a kinematic signature of the Galactic warp presented here is a preliminary study that adopts an
exploratory approach, aimed at determining whether there is evidence in the \emph{Gaia} DR1 and/or in the
pre-\emph{Gaia} global astrometry of the New \emph{Hipparcos} Reduction. While unexpected, our finding that distant OB
stars do not evidence the kinematic signature of the warp is in keeping with the previous results of \cite{Smart:1998}
and \cite{Drimmel:2000}, who analyzed the original \emph{Hipparcos} proper motions using a simpler approach than
employed here.

We point out that this work only considers a small fraction of the \emph{Gaia} DR1 data with a full astrometric
solution, being restricted to a subset of Hipparcos stars in DR1 brighter than $m_{V_T} = 8.5$. \emph{Gaia} DR1 TGAS
astrometry is complete to about $m_{V_T} = 11$, and potentially will permit us to sample a significantly larger volume
of the disk of the Milky Way than presented here. In future work we will expand our sample to a fainter magnitude
limit, using selection criteria based on multi-waveband photometry from other catalogues. We will also compare the
kinematics of this young population to an older population representative of the relaxed stellar disk.

Understanding the dynamical nature of the Galactic warp will need studies of both its structural form as well as
its associated kinematics. \emph{Gaia} was constructed to reveal the dynamics of the Milky Way on a large scale, and
we can only look forward to the future \emph{Gaia} data releases that will eventually contain astrometry for over a
billion stars. We expect that \emph{Gaia} will allow us to fully characterize the dynamical properties of the warp, as
suggested by \cite{Abedi:2014,Abedi:2015}, and allow us to arrive at a clearer understanding of the nature and origin
of the warp.  At the same time, \emph{Gaia} may possibly reveal other phenomenon causing systematic vertical velocities
in the disk of the Milky Way.


\begin{acknowledgements}
This work has been partially funded by MIUR, through PRIN 2012 grant No. 1.05.01.97.02 \textquotedbl Chemical and
dynamical evolution of our Galaxy and of the galaxies of the Local Group\textquotedbl\, and by ASI under contract No.
2014-025-R.1.2015 \textquotedbl \emph{Gaia} Mission - The Italian Participation to DPAC\textquotedbl.  E. Poggio acknowledges
the financial support of the 2014 PhD fellowship programme of INAF. This work has made use of data from the European
Space Agency (ESA) mission {\it Gaia} (http://www.cosmos.esa.int/gaia), processed by the {\it Gaia} Data Processing and
Analysis Consortium (DPAC, http://www.cosmos.esa.int/web/gaia/dpac/consortium). Funding for the DPAC has been provided
by national institutions, in particular the institutions participating in the {\it Gaia} Multilateral Agreement. The
authors thank the referee for comments and suggestions that improved the overall quality of the paper.
They also thank Scilla degl'Innocenti for the helpful discussions.
\end{acknowledgements}


\bibliography{mybib}{}
\bibliographystyle{aa}


\begin{appendix}

\section{Hipparcos astrometric errors}
The tables \ref{table:erralpha2007}, \ref{table:errdelta2007}, \ref{table:errplx2007}, \ref{tab:errmua2007} and \ref{table:errmud2007} show the median formal errors of right ascension $\alpha$, declination $\delta$, parallax $\varpi$, proper motion components $\mu_{\alpha}$ and $\mu_{\delta}$ in function of apparent magnitude and ecliptic latitude for the HIP2 stars. They were obtained considering the entire HIP2 catalogue \citep[as given by][]{VanLeeuwen2007} excluding the stars redder than $(B-V)=0.5$. We also excluded stars for which there was a claim of binarity in \cite{VanLeeuwen2007}, taking account for binary systems after the single stars errors are generated, as described in the text. To construct the tables, we binned the resulting sample of 15197 HIP2 stars with respect to apparent magnitude and ecliptic latitude and found the median errors for each bin. Table \ref{table:numerr2007} shows the number of objects in each bin.

\begin{table}[h]
  \caption{Median formal uncertainties for $\sigma_{\alpha}$.}
  \label{table:erralpha2007}
  \centering
  \begin{tabular}{cccccccc}
    \hline
    \hline
       & \multicolumn{7}{c}{Ecliptic latitude ($|\beta|$, (deg)) } \\
     $m_V$& 0-10 & 10-20 & 20-30 & 30-40 & 40-50 & 50-60 & 60-90 \\
    \hline
     3-4 &  0.22  & 0.19  & 0.40  & 0.23  & 0.12  & 0.11  & 0.20\\
     4-5 & 0.24  & 0.25  & 0.25  & 0.20  & 0.15  & 0.16  & 0.15\\
     5-6 & 0.34  & 0.32  & 0.31  & 0.27  & 0.21  & 0.19  & 0.20\\
     6-7 & 0.49  & 0.46  & 0.44  & 0.37  & 0.29  & 0.29  & 0.29\\
     7-8 & 0.66  & 0.64  & 0.60  & 0.51  & 0.40  & 0.39  & 0.40\\
     \hline
  \end{tabular}
\end{table}

\begin{table}[h]
  \caption{Median formal uncertainties for $\sigma_{\delta}$.}
  \label{table:errdelta2007}
  \centering
  \begin{tabular}{cccccccc}
    \hline
    \hline
       & \multicolumn{7}{c}{Ecliptic latitude ($|\beta|$, (deg)) } \\
     $m_V$& 0-10 & 10-20 & 20-30 & 30-40 & 40-50 & 50-60 & 60-90 \\
    \hline
     3-4 & 0.16  & 0.12  & 0.31  & 0.19  & 0.12  & 0.13  & 0.26 \\
     4-5 & 0.16  & 0.18  & 0.18  & 0.17  & 0.17  & 0.17  & 0.16 \\
     5-6 & 0.23  & 0.22  & 0.23  & 0.24  & 0.23  & 0.21  & 0.21 \\
     6-7 & 0.32  & 0.32  & 0.33  & 0.32  & 0.32  & 0.31  & 0.29 \\
     7-8 & 0.44  & 0.44  & 0.44  & 0.44  & 0.45  & 0.42  & 0.40\\
     \hline
  \end{tabular}
\end{table}

\begin{table}[h]
  \caption{Median formal uncertainties for $\sigma_{\varpi}$.}
  \label{table:errplx2007}
  \centering
  \begin{tabular}{cccccccc}
    \hline
    \hline
       & \multicolumn{7}{c}{Ecliptic latitude ($|\beta|$, (deg)) } \\
     $m_V$& 0-10 & 10-20 & 20-30 & 30-40 & 40-50 & 50-60 & 60-90 \\
    \hline
     3-4 &  0.20 & 0.19 & 0.19 & 0.21 & 0.16 & 0.13 & 0.14\\
     4-5 &  0.25 & 0.26 & 0.25 & 0.23 & 0.24 & 0.19 & 0.16 \\
     5-6 & 0.36 & 0.34 & 0.35 & 0.34 & 0.33 & 0.26 & 0.22 \\
     6-7 & 0.51 & 0.51 & 0.49 & 0.47 & 0.47 & 0.38 & 0.31 \\
     7-8 & 0.71 & 0.70 & 0.69 & 0.66 & 0.66 & 0.53 & 0.45 \\
     \hline
  \end{tabular}
\end{table}

\begin{table}[h]
\caption{Median formal uncertainties for $\sigma_{\mu_{\alpha*}}$.}
\label{tab:errmua2007}
\centering
  \begin{tabular}{cccccccc}
    \hline
    \hline
       & \multicolumn{7}{c}{Ecliptic latitude ($|\beta|$, (deg)) } \\
     $m_V$& 0-10 & 10-20 & 20-30 & 30-40 & 40-50 & 50-60 & 60-90 \\
    \hline
     3- 4 & 0.24 & 0.22 & 0.18 & 0.19 & 0.13 & 0.12 & 0.15 \\
     4-5 & 0.28 & 0.29 & 0.26 & 0.21 & 0.17 & 0.17 & 0.16 \\
     5-6 & 0.41 & 0.37 & 0.35 & 0.30 & 0.24 & 0.22 & 0.23 \\
     6-7 & 0.58 & 0.55 & 0.50 & 0.43 & 0.33 & 0.32 & 0.32 \\
     7-8 & 0.80 & 0.77 & 0.70 & 0.60 & 0.47 & 0.44 & 0.45 \\
     \hline
  \end{tabular}
\end{table}

\begin{table}[h]
 \caption{Median formal uncertainties for $\sigma_{\mu_{\delta}}$.}
  \label{table:errmud2007}
  \centering
  \begin{tabular}{cccccccc}
    \hline
    \hline
       & \multicolumn{7}{c}{Ecliptic latitude ($|\beta|$, (deg)) } \\
     $m_V$& 0-10 & 10-20 & 20-30 & 30-40 & 40-50 & 50-60 & 60-90 \\
    \hline
     3-4 &  0.16 & 0.16 & 0.16 & 0.18 & 0.13 & 0.13 & 0.15\\
     4-5 &  0.19 & 0.20 & 0.19 & 0.17 & 0.18 & 0.17 & 0.16 \\
     5-6 & 0.27 & 0.27 & 0.26 & 0.26 & 0.26 & 0.24 & 0.23 \\
     6-7 & 0.37 & 0.38 & 0.38 & 0.36 & 0.36 & 0.34 & 0.33 \\
     7-8 & 0.53 & 0.53 & 0.52 & 0.52 & 0.52 & 0.48 & 0.46 \\
     \hline
  \end{tabular}
\end{table}

\begin{table}[h]
  \caption{Number of stars in each bin.}
  \label{table:numerr2007}
  \centering
  \begin{tabular}{cccccccc}
    \hline
    \hline
       & \multicolumn{7}{c}{Ecliptic latitude ($|\beta|$, (deg)) } \\
     $m_V$& 0-10 & 10-20 & 20-30 & 30-40 & 40-50 & 50-60 & 60-90 \\
    \hline
     3-4 & 20 & 20 & 20 & 12 & 17 & 15 & 18\\
     4-5 &  69 & 65 & 63 & 57 & 61 & 51 & 61\\
     5-6 & 209 & 193 & 209 & 169 & 167 & 183 & 196\\
     6-7 &  565 & 550 & 581 & 538 & 522 & 514 & 577\\
     7-8 &   1274 & 1287 & 1416 & 1372 & 1355 & 1321 & 1447\\
     \hline
  \end{tabular}
\end{table}

\end{appendix}

\end{document}